\documentclass[journal=jacsat,manuscript=article]{achemso}
\usepackage[utf8]{inputenc}
\usepackage[T1]{fontenc}
\usepackage[english]{babel}
\usepackage{graphicx}
\usepackage{epsfig}
\usepackage{epstopdf}
\usepackage{amsmath}
\usepackage{amssymb}
\usepackage{times}
\usepackage{setspace}
\usepackage{verbatim}
\usepackage{color}
\usepackage{mathrsfs}
\usepackage{euscript}
\usepackage{mathptmx}
\usepackage{float}

\usepackage{hyperref}
\hypersetup{
    colorlinks=true,
    linkcolor=blue,
    filecolor=magenta,      
    urlcolor=black,%red,
    citecolor=red, 
    pdftitle={Lane},
    }

\author{Debarshee Bagchi}
\affiliation{International Centre for Theoretical Sciences, Tata Institute of Fundamental Research, Bengaluru, India}
\email{debarshee.bagchi@icts.res.in}

\title[Lane]
  {Macroscopic charge segregation in driven polyelectrolyte solutions}


\begin{document}
%
% \begin{tocentry}
% %\begin{center}
% {\includegraphics[width=8cm]{TOC.pdf}} 
% %\end{center}
% \end{tocentry}

\begin{abstract}
Understanding the behavior of charged complex fluids is crucial for a plethora of important industrial, technological, and medical applications. Using coarse-grained molecular dynamics simulations, here we investigate the properties of a polyelectrolyte solution, with explicit counterions and implicit solvent, that is driven by a steady electric field. By properly tuning the interplay between interparticle electrostatics and the applied electric field, we uncover two nonequilibrium continuous phase transitions as a function of the driving field. The first transition occurs from a homogeneously mixed phase to a macroscopically charge segregated phase, in which the polyelectrolyte solution self-organizes to form two lanes of like-charges, parallel to the applied field. We show that the fundamental underlying factor responsible for the emergence of this charge segregation in the presence of electric field is the excluded volume interactions of the drifting polyelectrolyte chains. As the drive is increased further, a re-entrant transition is observed from a charge segregated phase to a homogeneous phase. The re-entrance is signaled by the decrease in mobility of the monomers and counterions, as the electric field is increased. Furthermore, with multivalent counterions, a counterintuitive regime of negative differential mobility is observed, in which the charges move progressively slower as the driving field is increased. We show that all these features can be consistently explained by an intuitive trapping mechanism that operates between the oppositely moving charges, and present numerical evidence to support our claims. Parameter dependencies and phase diagrams are studied to better understand charge segregation in such driven polyelectrolyte solutions.
\end{abstract}
%\pacs{}
%\maketitle

\section{Introduction}
Polyelectrolytes are polymers whose monomers contain ionizable groups and are ubiquitous in biological and industrial systems \cite{dobrynin2005theory,netz2007polyelectrolytes,rubinstein2012polyelectrolytes,muthukumar201750th}. Examples of naturally occurring polyelectrolytes are DNA, RNA, and proteins, but nowadays there are also widely used synthetic polyelectrolytes, such as polystyrene sulfonate and polyacrylic acid. When placed in a polar solvent, such as water, polyelectrolyte monomers ionize, thereby acquiring charge, and release oppositely charged counterions into the solvent. Due to the complex interplay among conformational entropic effects, excluded volume interactions, and electrostatics, polyelectrolytes often exhibit structural and dynamical properties that are drastically different from electrolytes, colloids, and neutral polymers. The unique properties of polyelectrolytes are governed by a multitude of nonlinearly coupled variables, making these systems notoriously difficult to study, both theoretically and in experiments. Needless to say, due to their novel features, polyelectrolyte systems have contributed immensely, both in understanding the fundamental physics of complex charged fluids, and for a wide range of real life applications, such as nano-filtration \cite{jin2003use}, energy storage \cite{wang2011self}, bioelectronics \cite{kim2020dna}, fabrication of novel materials for biomedical use \cite{ishihara2019polyelectrolyte}, and vaccines \cite{mulligan2020phase}, to name a few.

In this work, we study the nonequilibrium aspects of an aqueous polyelectrolyte solution, driven by an electric field, using coarse-grained simulations. Electrokinetically driven polyelectrolytes have been actively investigated over the years in a variety of different contexts \cite{netz2003nonequilibrium,netz2003polyelectrolytes,brun2008dynamics,wei2009unfolding,tang2011compression,klotz2017dynamics,suma2018electric,mahinthichaichan2020polyelectrolyte,bagchi2020dynamics,radhakrishnan2021collapse}. This also has important applications, such as in the electrophoretic separation of DNA and other polyelectrolytes \cite{viovy2000electrophoresis} which constitute an useful way to probe and characterize their structure and properties. Another of our major motivations for investigating this problem stems from the previous studies of driven colloidal systems, mostly done in two dimensions (2D), and a few in three dimensions (3D). For 2D and 3D binary colloids driven by an external field, a nonequilibrium phase transition has been reported  which is sometimes referred to as the {\it laning transition}  \cite{dzubiella2002lane,rex2007lane,rex2008lane,frank2009mesoscale,wysocki2009oscillatory,glanz2012nature,li2021phase}. Molecular simulations show that, in the presence of an appreciably large electric field, a colloidal binary mixture spontaneously forms multiple lanes of like-charges. This phenomenon has been explained in terms of dynamical instability using density functional theory \cite{chakrabarti2003dynamical,chakrabarti2004reentrance}, nonequilibrium correlation functions derived from many-body Smoluchowski equation \cite{kohl2012microscopic}, environment dependent diffusion rates and analogy with the Katz-Lebowitz-Spohn model \cite{klymko2016microscopic}, and superadiabatic demixing forces \cite{geigenfeind2020superadiabatic}. Lane formation has also been verified in experiments with oppositely charged colloids \cite{leunissen2005ionic,vissers2011lane}. 
For 2D systems, lane formation is believed to be a discontinuous transition, or possibly a smooth crossover. For 3D charged systems, it is expected to be a continuous transition. Nonetheless, the true nature of laning transition in binary colloids is still debated \cite{glanz2012nature,klymko2016microscopic,poncet2017universal}. This is mostly because, unlike equilibrium, it is often extremely nontrivial to identify and characterize nonequilibrium phase transitions in complex systems. This also makes nonequilibrium transitions exceedingly appealing to study and has important consequences, e.g., in understanding self-assembly processes in soft-matter systems \cite{grunwald2016exploiting}.

To examine the generality of laning, it seems worthwhile to investigate if such a nonequilibrium phase transition occurs also in other classes of complex fluids, such as polyelectrolyte solutions. We speculate that, if laning transition exists for polyelectrolyte systems, then there should be considerable quantitative, as well as, qualitative differences in its features (compared to colloids), due to the complex interplay among excluded volume interactions of the spatially extended polyelectrolyte chains, entropic effects, and the electrostatic interactions. To the best of our knowledge, laning transition in driven polyelectrolyte solutions has not been studied yet, and we attempt to address this question in this article.

To this end, we perform nonequilibrium Langevin dynamics simulations of coarse-grained bead-spring polyelectrolytes, with explicit counterions in an implicit water-like solvent, that is driven by a steady external electric field. By judiciously controlling the strength of the electrostatic interaction between the charged species and its interplay with the electric field drive, our simulations unravel a {\it macroscopic nonequilibrium charge segregation} between the monomers and the oppositely charged counterions, a {\it re-entrant phase transition} at higher electric fields, and another associated transition between regimes of increasing and decreasing mobility with the electric field drive. We also find an anomalous regime of {\it negative differential mobility}, in which, remarkably, the terminal velocity of the charged particles decreases as the driving field strength is increased. In the following sections, we describe our model and methods in detail, present the results obtained from simulations, and provide simple physical arguments to elucidate the nonequilibrium behavior of driven polyelectrolytes.
% 
% The remainder of the paper is organized as follows. In Sec. \ref{model}, we describe our coarse-grained polyelectrolyte model, the Langevin dynamics simulations, and the important observables that have been computed to characterize the nonequilibrium properties of the system. The simulation results are presented and explained in Sec. \ref{results}. We conclude with a summary of the main results and future outlook in Sec. \ref{conclusions}.
% 

\section{Model and simulation method}
\label{model}
\paragraph{Model.}
We consider polyelectrolytes that are modeled as linear, fully flexible, bead-spring chains, with bead diameter $\sigma$ and mass set to unity (for a recent review on coarse-grained molecular simulations of charged polymers, see Ref. \cite{shen2021molecular}). There are $N_p$ polyelectrolyte chains, each with $M_p$ monomers. Each monomer carries one unit of negative charge $-q$. To ensure global charge neutrality, there are $N_c = N_p \times M_p$ counterions (diameter $\sigma$ and unit mass), each carrying one unit of positive charge $+q$.
Excluded volume interaction between the beads is modeled using the purely repulsive Lennard-Jones potential, also referred to as the Weeks-Chandler-Andersen (WCA) potential, 
$$V_{WCA}(r) = 4\epsilon[(\sigma/r)^{12} - (\sigma/r)^{6}],$$
with a cutoff $r_c = 2^{1/6}\sigma$, such that $V_{WCA}(r > r_c) = 0$. All quantities here are measured in reduced units, where $\sigma$ and $\epsilon$ are the length and energy scales respectively.
The consecutive monomers on a polyelectrolyte chain are connected by the finite extensible nonlinear elastic (FENE) potential,
$$V_b(r) = - \dfrac 12 k R_0^2 \left(1 - \dfrac{r^2}{R_0^2}\right),$$ 
where $k = 30$ is the spring constant, and $R_0 = 1.5$ is the maximum extension of the bonds. 
Following the studies on colloidal systems, the charges in our system (monomers and counterions) interact via the screened Coulomb interaction (also known as the Yukawa or the Debye-H\"{u}ckel potential). The screened Coulomb interaction, besides being computationally (and analytically) easier to handle than the full long-range Coulomb interaction, is usually found in many biologically and industrially relevant systems, due to the presence of dissolved salt ions in the aqueous solution \cite{dobrynin2005theory,netz2007polyelectrolytes,rubinstein2012polyelectrolytes,muthukumar201750th}. It has the exponentially decaying form
$$V_q(r) = k_BT  {\ell}_B \dfrac{e^{-\kappa r}}r,$$ 
and we choose the electrostatic cutoff at $r_q = r_c$. Here $\kappa$ %= (8\pi{\ell_B}c_s)^{1/2}$%
is the inverse Debye screening length, and ${\ell_B}$ is the Bjerrum length.
The solvent is implicit, and therefore, is assumed to be a continuum neutral background with no internal structure. We choose the solvent to be water which has a dielectric constant $\varepsilon_r = 80$, and Bjerrum length ${\ell}_B = \frac{q^2}{\varepsilon_r k_BT} \approx 0.7$ nm $= 0.7 \sigma$. The solvent also provides dissipation and thermal noise, and is maintained at a temperature $T$, with friction coefficient always set to $\gamma = 1$. This gives us a simple coarse-grained model for polyelectrolytes, with explicit counterions in an implicit solvent, and screened Coulomb interactions between the charges.

% $$\frac{d^2r_i}{dt^2} = F_{WCA,i} + F_{q,i} + F_{b,i} - \gamma \frac{dr_i}{dt} + \eta_i$$

The polyelectrolyte solution is placed in a 3D periodic cubic simulation box of volume $L^3$, where $L$ is chosen according to the desired volume fraction $\rho =\frac{N \pi \sigma^3}{6L^3}$, where $N = N_p \times M_p + N_c$ is the total number of beads inside the simulation box. In this work we have considered both monovalent and multivalent (trivalent) counterions. For trivalent counterions, the charge on each counterion is increased to $3q$, whereas their number is reduced to $N_c = \frac 13 (N_p \times M_p)$, to maintain global charge neutrality. This polyelectrolyte solution is driven by a steady external electric field ${\bf E} \equiv \{0,0,E\}$, directed toward to the $+\hat z$ direction. Thus, when the external field is switched on, the polyelectrolyte chains tend to move toward the $-\hat z$ and the counterions move in the $+\hat z$ direction.

\paragraph{Simulation method.}
Starting from random positions and velocities for the monomers and the counterions, the polyelectrolyte solution is simulated using Langevin dynamics in LAMMPS \cite{plimpton1995fast}. The update time-step is set at $\Delta t = 0.001$. First, the polyelectrolyte solution is equilibrated at temperature $T$ without the external drive, and thereafter, the electric field is turned on. In the presence of driving, the system is allowed to settle to a nonequilibrium steady state (NESS). At NESS, we compute average values of different observable (defined below), such as the order parameter $\Psi({\bf E})$ and the transport coefficients i.e., the mobilities $\mu_{p,c}({\bf E})$ -- the subscripts $p$ and $c$ stand for the polyelectrolytes and the counterions respectively.

\paragraph{Observables.}
To characterize the phase transition a suitable order parameter needs to be defined. The order parameter is non-unique and can be defined in multiple ways. For this work we define an order parameter, well-suited for our purpose, in the following way. A relevant length scale that needs to be considered is the mean inter-particle distance $a = \left(L^3/N\right)^{1/3}$. Since charge segregation happens perpendicular to the $x-y$ plane, we first compute the number of counterions $n_c$ that are at an arbitrarily chosen distance $\delta = \sqrt{(x_i - x_j)^2 + (y_i - y_j)^2} < \frac a2$ from a monomer on any of the polyelectrolyte chains \cite{dzubiella2002lane}. Here, $\{x_i, y_i\}$ and $\{x_j, y_j\}$ are the $x-y$ coordinates of the $i-$th counterion and the $j-$th monomer respectively. We then define the order parameter as $\Psi = 1 - \frac{n_c}{N_c}$, such that in the homogeneous state $n_c \approx N_c$, and hence, $\Psi \approx 0$, whereas, in the charge segregated phase, $n_c << N_c$, and $\Psi$ has a large value close to unity. The relevant transport coefficient in this problem is the mobility of the charges in the presence of external drive. The mobility $\mu$ is generally defined as the average drift velocity $<v_z>$ of a monomer/counterion in the NESS, normalized by the magnitude of the force due to the applied field $F_q = \pm qE$ (`$+$' for counterions and `$-$' for monomers), i.e., $\mu = \frac{<v_z>}{F_q}$, where $<...>$ implies average over all the beads of the same species and time.

\section{Results and discussions}
\label{results}
\begin{figure*}[tbh]
\centering
{\includegraphics[width=16cm]{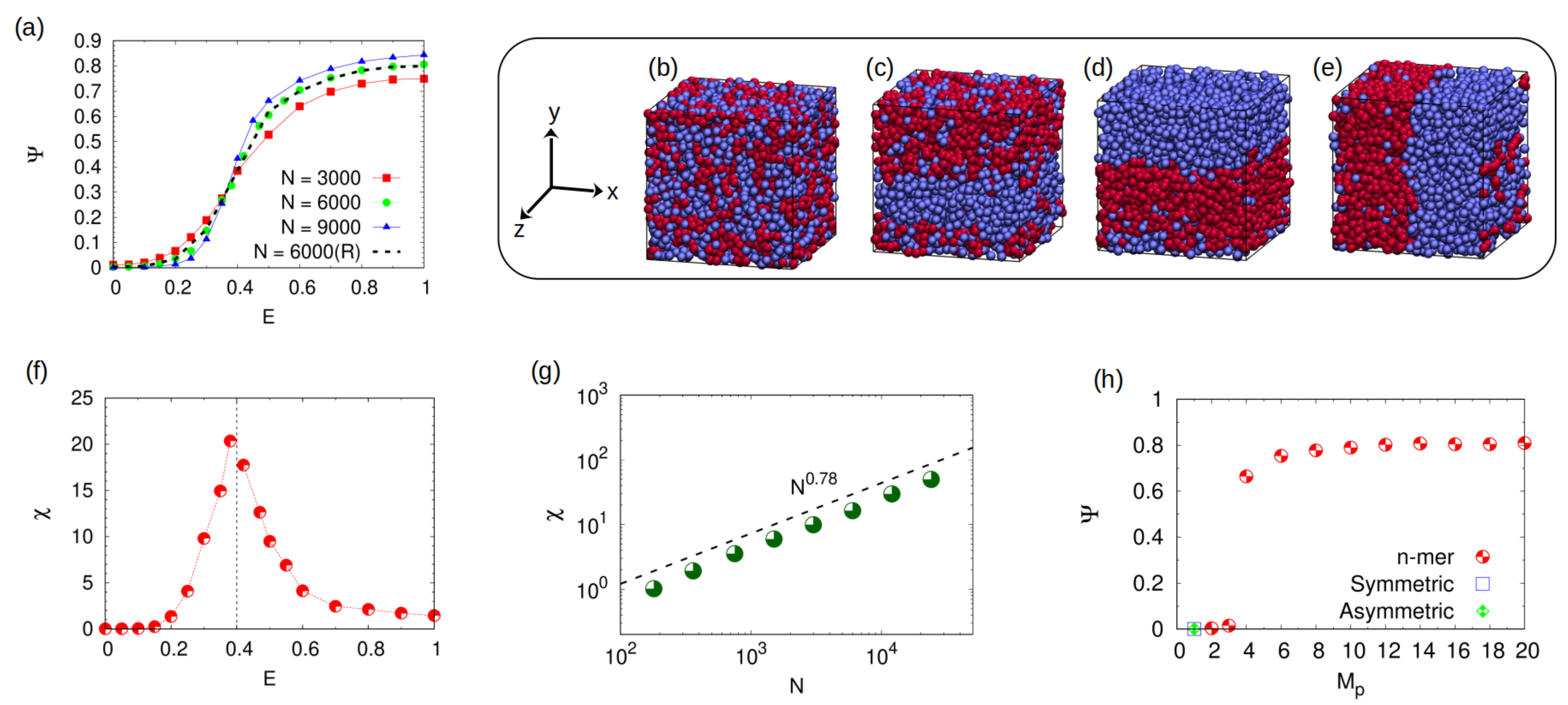}}
\caption{(a) Order parameter $\Psi(E)$ with applied electric field E, for $N = 3000, 6000$, and $9000$. Density $\rho = 0.25$ is kept fixed. The dashed line, labeled `$N=6000(R)$', shows the order parameter starting from a fully charge segregated initial configuration. In (b)--(e), simulation snapshots are shown for $N = 6000$ and $E = 0, 0.3, 0.7$, and $1$ respectively. Negatively charged monomers are represented by red beads and positively charged counterions are represented by blue beads. Periodic boundary conditions are imposed along all three directions. (f) Susceptibility $\chi$ corresponding to $\Psi(E)$ in (a) for $N = 6000$. (g) Power-law divergence of $\chi(N)$ near the transition point $E \approx 0.4$. (h) Order parameter of $n$-mers, charge symmetric electrolytes, and charge asymmetric electrolytes, for $E = 1.0$, $N = 6000$, and $\rho = 0.25$.}
\label{fig:Fig1}
\end{figure*}

\paragraph{Macroscopic charge segregation.}
We first show that the polyelectrolyte solution undergoes a nonequilibrium phase transition, from a macroscopically homogeneous mixture of negatively charged monomers and positively charged counterions to a macroscopically charge segregated phase, in the presence of the applied electric field. This can be seen in Fig. \ref{fig:Fig1}a, where the order parameter $\Psi$ is shown to increase smoothly from zero to a nonzero value (close to unity) as $E$ is increased, indicating charge segregation. Moreover, the phase transition becomes sharper as the number of charges, $N$, is increased, suggesting that the transition persists in the thermodynamic limit. Thus, at large $E$, quite counter-intuitively, the positively charged counterions and the negatively charged monomers diffuse {\it uphill} \cite{killer2016phase}, and spontaneously organize to form two separate lanes of like-charges, despite the electrostatic repulsion between the charges in the same lane. If the driving field is turned off the polyelectrolyte solution becomes homogeneous again. A few representative simulation snapshots of the polyelectrolyte solution for different electric fields are shown in Figs. \ref{fig:Fig1}b-e clearly showing the phase segregation. The symmetry breaking can happen spontaneously along any of the two transverse directions, $\hat x$ or $\hat y$, see Figs. \ref{fig:Fig1}d and \ref{fig:Fig1}e (depending on the density and external drive, other demixed configurations are also possible, see Fig. S1 in the \textcolor{blue}{Supporting Information}).

Note that, unlike 2D and 3D colloidal systems \cite{dzubiella2002lane}, in our  polyelectrolyte system (3D), we do not obtain multiple lanes, but only two macroscopic lanes. It has been speculated that, even for 2D colloids, eventually the multiple lanes (of the same charge) will merge, leading to two macroscopic lanes \cite{dzubiella2002lane}, just as in our case. However, starting from a completely random initial configuration of charges, it will take infinitely long to obtain two macroscopic lanes in 2D. This is because, unlike 3D systems where there is a third direction available, in 2D the individual lanes of like-charges have to pass through nearby lanes of opposite charges in order to merge. In fact, starting from two macroscopically segregated phases, it has been demonstrated that a 2D system maintains the two lanes in the NESS (see Fig. 9b in Ref. \cite{klymko2016microscopic}).

\paragraph{Hysteresis and susceptibility.}
To understand the nature of the nonequilibrium phase transition, we look at the variation of $\Psi(E)$ as the external drive is gradually increased from $E=0$ to $E=1$, and thereafter decreased back to zero. The order parameter along the forward path (marked `N = 6000' in Fig. \ref{fig:Fig1}a) and the reverse path (dashed curve marked `N = 6000(R)') does not exhibit any noticeable hysteresis, which can be taken as a signature of a continuous phase transition, in analogy with equilibrium critical phenomena. This is also the case for 3D colloidal systems, but not for 2D colloids \cite{rex2008lane,vissers2011lane}.
We also compute the susceptibility $\chi =  <\Psi^2> - <\Psi>^2$ (fluctuations of the order parameter), see Fig. \ref{fig:Fig1}f. Near the transition point, $E = E_c \approx 0.4$, $\chi(E)$ has a relatively high value, implying that there are large critical fluctuations. Moreover, near the transition point, $\chi$ increases as a power-law with $N$, $\chi \sim N^{0.78}$, for roughly two decades (Fig. \ref{fig:Fig1}g), as is also expected in equilibrium critical phenomena.

\paragraph{Electrolytes and n--mers.}
To demonstrate that this phase transition, for the chosen system parameters and external $E$-field range, is unique to the polyelectrolyte system, we simulated an equivalent symmetric electrolyte solution. The equivalent electrolyte solution is identical to the polyelectrolyte solution, except that all the bond potentials are set to zero, i.e., $V_b(r) = 0$. We fix the external electric field to $E = 1.0$ which induces a strong segregation in the polyelectrolyte solution (Fig. \ref{fig:Fig1}e). It can be clearly seen in Fig. \ref{fig:Fig1}h that, at such a low external drive, the symmetric electrolyte solution, containing $N_+$ positive ions and $N_-$ negative ions, does not phase segregate ($\Psi \approx 0$). (See Ref. \cite{netz2003conduction} for lane formation in 2D electrolytes at very high $E$-field.)

Building up from the symmetric $1:1$ electrolyte, and keeping $N$ and $\rho$ fixed, we increase the degree of polymerization of the chains, $M_p$. Thus $M_p = 2$ is a dimer, and $M_p = 3$ is a trimer, and in general $M_p = n$ is an `{\it n}--mer'. For a fixed drive $E=1.0$, we find that the order parameter $\Psi$, indicating phase segregation, increases with $M_p$. For $M_p \geq 4$, $\Psi$ has a large finite value, and it becomes almost constant for $M_p \gtrsim 10$. Unless mentioned otherwise, we set $M_p = 15$ for the simulation results presented in this article. For the same electric field range, we have also verified that such a phase transition is absent even in a charge asymmetric electrolyte solution, where the $N_+$ positive ions carry $+q$ charge each, the $N_-$ negative ions carry $-4q$ charge each, and $N_+ = 4N_-$ (Fig. \ref{fig:Fig1}h). Typical NESS snapshots for all these cases are provided in Fig. S2 in the \textcolor{blue}{Supporting Information}.

\paragraph{Explanation of charge segregation.}
Based on the results described above, we present a simple physical explanation for the emergence of charge segregation in the polyelectrolyte solution. In the presence of the external drive, the oppositely moving polyelectrolyte chains and counterions collide frequently with each other. As the electric field is increased, the two oppositely charged species start to move increasingly faster, producing more and more collisions between them. In order to avoid the energetically expensive collisions (friction) between the sliding charge fronts of polyelectrolytes and counterions, 
the colliding charges tend to diffuse away from each other along the transverse directions to the drive ($\hat x$ and $\hat y$ in our case). Consequently, the polyelectrolyte solution self-organizes and segregates to form distinct lanes of like-charges, thereby enhancing transport. Moreover, since chains with $M_p \geq 4$ show segregation but not the charge symmetric/asymmetric electrolytes with the same amount of total charge (Fig. \ref{fig:Fig1}h), this suggests that it is the excluded volume effects of the spatially extended flexible polyelectrolyte chains, and the resulting friction, that trigger charge segregation (lane formation) in the polyelectrolyte solution.
\begin{figure}[htb]
\centering
{\includegraphics[width=8cm]{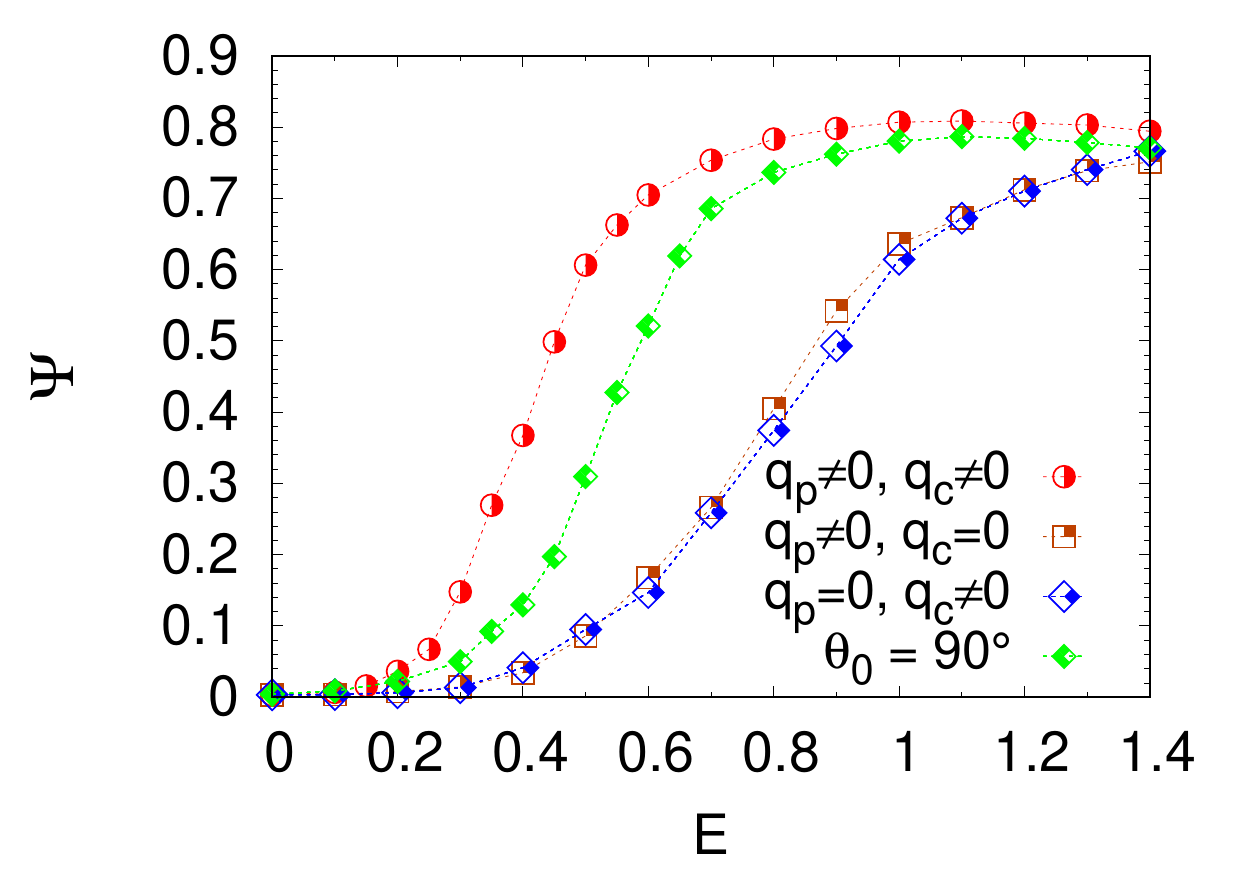}}%\hskip-0.5cm
\caption{Order parameter $\Psi(E)$ for a few test cases compared with the order parameter of the original model. Here $q_p$ and $q_c$ represent the charge on the polyelectrolyte chains and the counterions respectively. Even when only one of the species is charged ($q_p = 0$ or $q_c = 0$), the solution phase segregates. The semi-flexible polyelectrolyte chains, with equilibrium bond angle $\theta_0 = 90^\circ$, shows a lower degree of segregation compared to fully flexible polyelectrolyte chains (half-shaded red circles). All the other simulation parameters are the same as in Fig. 1.}
\label{fig:Expl}
\end{figure}

Our explanation can be made more convincing by studying a few instructive test cases. First, we make one of the species, either the counterions or the polyelectrolyte chains, charge neutral. Even for this somewhat pathological test case, a macroscopic segregation is observed in the presence of the electric field drive. This is shown in Fig. \ref{fig:Expl}, where $q_p$ ($q_c$) denotes the charge on the polyelectrolyte chains (counterions). Note that phase segregation with only one charged species is less pronounced, since the relative motion between the charges is reduced, and therefore, the friction is lower, for a fixed drive.
For the second test case, we intentionally suppress the excluded volume interactions of the polyelectrolyte chains. This can be achieved in simulations by using bond angle potentials of the form $V_a(\theta) = \frac 12 k_a (\theta - \theta_0)^2$ with equilibrium angle $\theta_0 = 90^\circ$ between the polyelectrolyte bonds. The bond angles restrict the polyelectrolyte chains to preferentially more collapsed configurations, and consequently, suppress excluded volume interactions. In Fig. \ref{fig:Expl}, the comparison of $\Psi$ for the fully flexible chain (half-shaded red circles) and the semi-flexible chain (labeled as `$\theta_0 = 90^\circ$') is shown. Phase segregation is found to be suppressed and shifted to higher $E$-fields for polyelectrolyte chains with reduced excluded volume. Thus, these two additional test cases demonstrate convincingly that it is the relative motion, and not the electrostatic interactions between the two species, that is responsible for the emergence of charge segregation. The moving polyelectrolyte chains, due to their chain connectivity and the resulting excluded volume effects, `sweep away' the oppositely moving counterions from their path in order to avoid the energetically costly collisions, and the system spontaneously reorganizes to form two separate lanes of like-charges. Electrostatic interactions only affect the segregation phenomenon quantitatively, such as the critical electric field magnitude and the degree of phase segregation. Also, in the NESS, instead of multiple lanes, only two macroscopic lanes form (as seen in the snapshots of Figs. \ref{fig:Fig1}d and \ref{fig:Fig1}e), since such a charge configuration leads to minimum number of collision events between the two species (hence lowest friction), and is therefore maximally favored.

\begin{figure}[htb]
\centering
{\includegraphics[width=8.3cm]{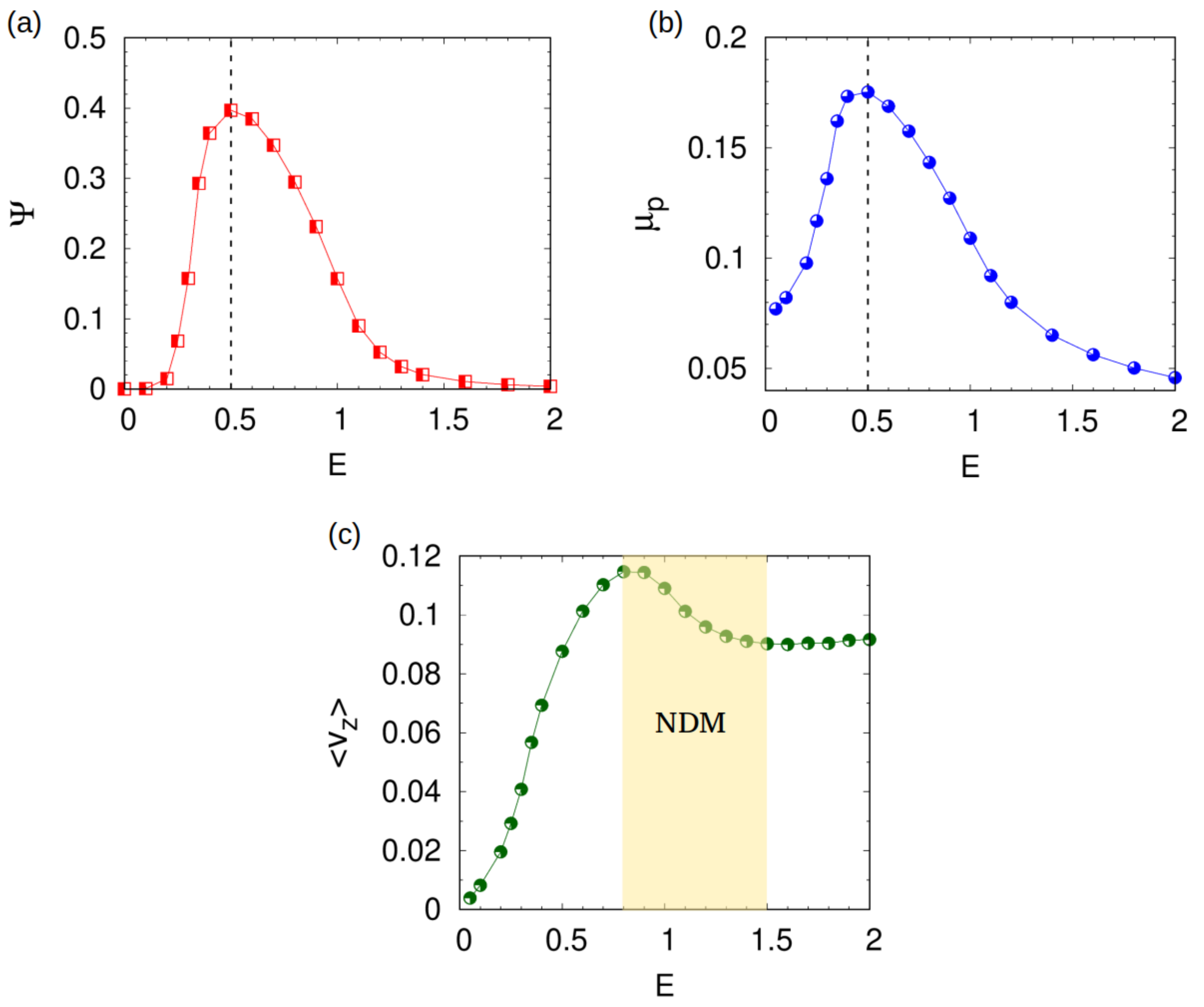}}%\hskip-0.5cm
\caption{(a) Order parameter $\Psi(E)$ for the polyelectrolyte solution with trivalent counterions. The re-entrance regime starts for $E \gtrsim 0.5$. For $E \gtrsim 1.5$, we have $\Psi \approx 0$, indicating a complete re-entrance to a homogeneous phase. (b) Mobility $\mu(E)$ decreases for $E \gtrsim 0.5$ indicating a negative slope $\frac{d\mu}{dE} < 0$. (c) The average drift velocity of the charged particles, $<v_z>$, is shown for different $E$-fields. The shaded region is the anomalous NDM regime, where the charges move slower with increasing external drive. Here, each of the $N_c = \frac 13 N_p \times M_P$ counterions carries $3q$ units of positive charge, and $N = 4000$. All the other parameters are the same as in Fig. \ref{fig:Fig1}.}
\label{fig:Fig2}
\end{figure}

\paragraph{Re-entrance phenomenon.}
Strikingly enough, keeping all the parameters the same, if we increase the electric field beyond $E = 1.0$ (in Fig. \ref{fig:Fig1}a), the order parameter $\Psi$ and the mobilities $\mu_{p,c}$ start to decrease (see Fig. S3 for data beyond $E = 1.0$ in the \textcolor{blue}{Supporting Information}). This implies that the macroscopically charge segregated system begins to approach a homogeneously mixed state at larger external drives. It seems reasonable to expect that, for extremely large electric fields, one would observe a total re-entrant transition from a strongly demixed state to a fully mixed state again, indicated by $\Psi \approx 0$. Since it is computationally difficult to access very large electric fields for a polyelectrolyte solution (due the presence of the FENE bonds), we speculate that the effects of strong nonlinearity would perhaps be more prominent and therefore easier to observe in the presence of multivalent counterions.

The segregation order parameter of the polyelectrolyte solution with trivalent counterions is shown in Fig. \ref{fig:Fig2}a. For this case, the re-entrance from a phase segregated to a mixed phase ($\Psi \approx 0$) can be clearly seen. Compared to monovalent counterions, the phase segregation with trivalent counterions is much less pronounced due to the stronger electrostatic correlations in the latter, and total re-entrance ($\Psi \approx 0$) happens at a much lower drive, $E \gtrsim 1.5$. Thus, for very small ($E \lesssim 0.25$) and large ($E \gtrsim 1.5$) external drives, we obtain $\Psi \approx 0$, whereas for $0.25 \lesssim E \lesssim 1.5$, we find charge segregation ($\Psi \neq 0$). We believe that the behavior for the monovalent counterion case will be qualitatively similar, but it will require much higher electric fields to observe the complete re-entrance. To the best of our knowledge, a re-entrance of this type has not been reported in the studies of driven binary colloids so far, and therefore, it remains to be seen, if such a phenomenon occurs there as well, for extremely large external drives. For 2D athermal disks with additional friction built into the model, a similar re-entrance behavior has been recently observed \cite{samsuzzaman2021reentrant}. For 2D colloids, a different kind of re-entrance has been reported in the past \cite{chakrabarti2004reentrance}, that happens for a fixed $E$ as the volume fraction $\rho$ is increased. Representative snapshots of the multivalent polyelectrolyte solution for different $E$-fields are provided in Fig. S4 of the \textcolor{blue}{Supporting Information}.

\paragraph{Mobilities.}
The signature of the phase segregation and re-entrance phenomena can also be clearly observed by computing the mobilities $\mu_p$ and $\mu_c$. The two mobilities show identical behavior in the NESS, and therefore, we drop the subscripts: $\mu_p = \mu_c = \mu$. As the external drive in increased from zero, the polyelectrolyte solution starts to phase segregate, and $\mu$ gradually increases. Remarkably, beyond $E = E_m \approx 0.5$ (Fig. \ref{fig:Fig2}a), when the electric field is increased further, $\mu$, instead of increasing, decreases sharply, see Fig. \ref{fig:Fig2}b. In other words, as the electric field is increased from zero, the mobility exhibits a transition from a positive slope ($\frac{d\mu}{dE} > 0$) to a negative slope ($\frac{d\mu}{dE} < 0$), which coincides perfectly with the emergence of the re-entrance phase. This reduction in mobility for $E > E_m$ reduces the friction between the oppositely moving charges and allows the system to again get mixed, leading to the re-entrance.

Furthermore, along the negative slope of the $\mu \sim E$ curve, for $0.8 \lesssim E \lesssim 1.5$, quite counter-intuitively, we find a regime where the average drift velocity of the charges, $<v_z>$, decreases as the external drive is increased. This anomalous phenomenon is known as the negative differential mobility (NDM), defined as $\mu_d = \frac{d<v_z>}{dE} < 0$ (shaded portion in Fig. \ref{fig:Fig2}c). Here again, we have omitted the subscripts, and normalized the drift velocities, $<v_{zp}>$ and $<v_{zc}>$, by their respective charges: $ -\frac 1q <v_{zp}> = \frac 1{3q} <v_{zc}> = <v_z>$. It is to be noted that NDM ($\mu_d < 0$) is not the same as absolute negative mobility, $\mu < 0$, which has also been observed in some systems \cite{eichhorn2002brownian,joanny2003motion,ros2005absolute,machura2007absolute}. 

The puzzling NDM regime, and negative response functions in general, have been observed and extensively studied in a variety of systems relevant to statistical physics, soft matter physics, and condensed matter physics. This includes minimalistic toy models of random walkers, exclusion processes, and conserved lattice gas models \cite{zia2002getting,basu2014mobility,benichou2014microscopic,baiesi2015role,chatterjee2018negative}, to more complicated and realistic physical systems, such as, colloidal particles \cite{eichhorn2010negative}, glass formers \cite{jack2008negative}, laminar flows \cite{sarracino2016nonlinear}, ring polymers in gel electrophoresis \cite{michieletto2015rings, iubini2018topological}, active matter systems \cite{reichhardt2017negative}, and superconductors \cite{gutierrez2009transition}, to name a few.
Other transport coefficients have also been known to show similar negative differential response far away from equilibrium, e.g., heat conductivity in the presence of a thermal bias \cite{li2006negative,yang2007thermal,bagchi2013thermal,bagchi2015thermally}. Thus, negative responses, described aptly by the phrase {\it `getting more from pushing less'} \cite{zia2002getting}, can be expected to appear quite generally under nonequilibrium conditions. Here we present another many-body interacting system, namely, linear polyelectrolytes with multivalent counterions, that exhibits NDM in the presence of an electric field drive.

Based on the ideas behind the formulation of generalized fluctuation-response relation in nonequilibrium \cite{zia2002getting, baiesi2009fluctuations,baerts2013frenetic, basu2014mobility}, we describe below an intuitive {\it trapping mechanism} to explain the re-entrance, the emergence of the negative slope in the $\mu \sim E$ curve, and the NDM regime. The explanation relies on the fact that at lower electric fields, $E < E_m$, when polyelectrolyte chains encounter the oppositely moving counterions, the two species can easily diffuse along the transverse directions ($\hat x, \hat y$), and glide past each other without getting strongly trapped (or stuck) for very long. However, as the electric field is increased to values beyond $E = E_m$, the probability to escape the trap becomes smaller. The polyelectrolyte chains and the counterions get strongly trapped with one another more frequently, and each tries to drag the opposite species along with them toward the {\it wrong} direction. 
% In other words, each species behaves as an oppositely moving obstacle (or trap) for the other species, and the {\it escape rate} to glide past the obstacle (or, to come out of the trap) is relatively low when the driving electric field is very high. 
%
Thus the polyelectrolyte chains are trapped and dragged by the counterions in the $+ \hat z$ direction, and vice-versa. This trapping mechanism creates hindrance to the unidirectional motion of the charges, thus slowing them down and leading to the decrease of the mobility with $E$, as seen in Fig. \ref{fig:Fig2}b for large $E > E_m$. For some values of the electric field, the effect of this {\it trap-and-drag} mechanism can become larger than electrokinetic push of the external drive, leading to the emergence of the counterintuitive NDM regime observed in Fig. \ref{fig:Fig2}c. 

A subtle point to note here is that, unlike $<v_z(E)>$, $<v_z^2(E)>$ is always monotonic, and hence the existence of NDM is not apparent from the variation of $<v_z^2>$ with $E$ (see Fig. S5 in the \textcolor{blue}{Supporting Information}). However $<v_z(E)>$ decreases in the NDM regime due to the rapid (undirected) back and forth motion of the charged particles along the $\hat z$ direction, often described by terms such as `nervosity' \cite{basu2014mobility}, `jitteriness' \cite{sarracino2016nonlinear} and `frenzy/frantic' \cite{maes2020frenesy} etc., due to the trap-and-drag effect produced by the charges moving in the opposite direction. The nonequilibrium fluctuation-response formula \cite{baiesi2009fluctuations,baerts2013frenetic} predicts that NDM may arise in general when this {\it frenetic term} becomes sufficiently negative compared to the usual entropic term of the generalized Green-Kubo formula.

\begin{figure}[htb]
\centering
{\includegraphics[width=8.65cm]{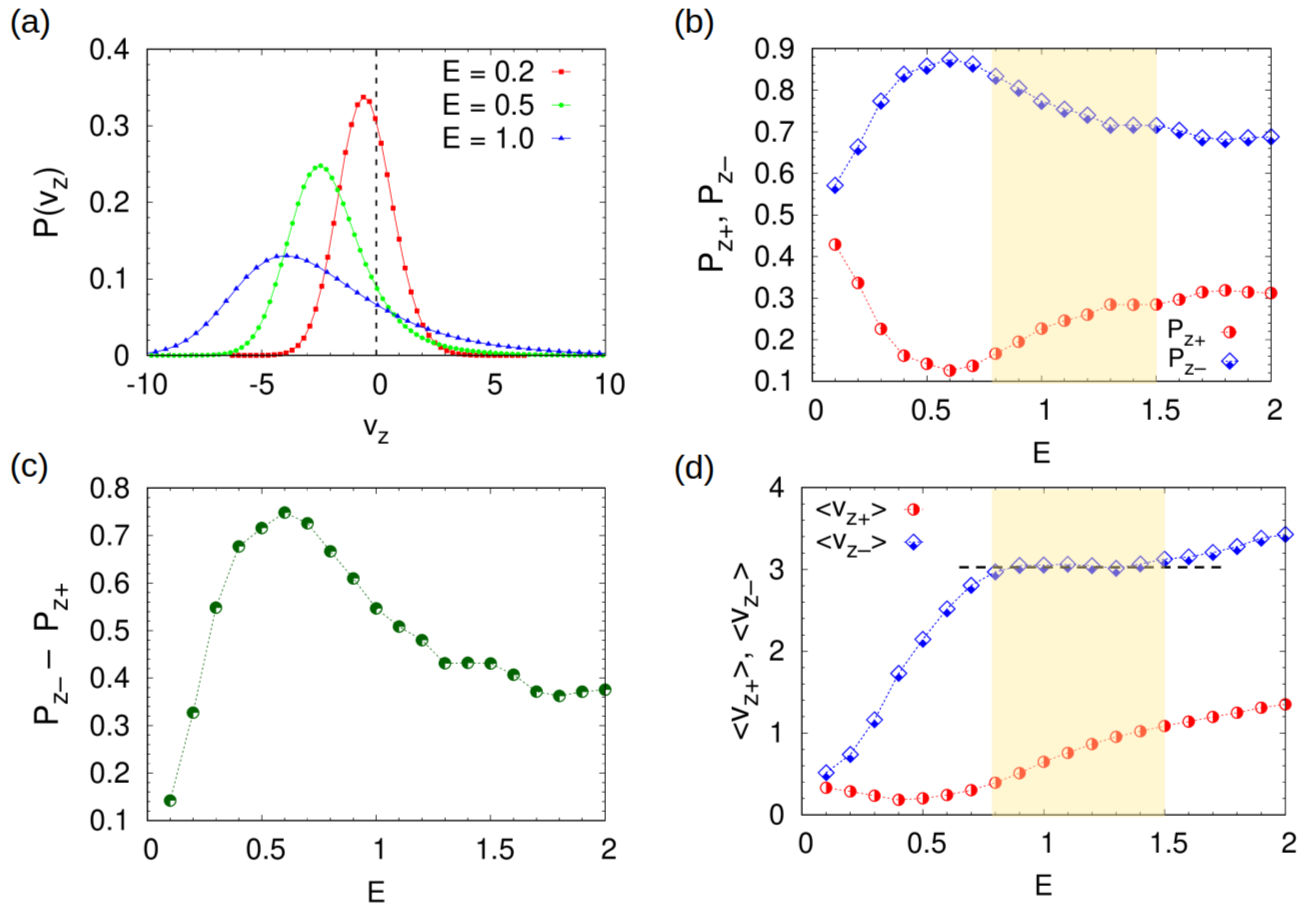}}%\hskip-0.5cm
\caption{(a) Distribution of drift velocity of the monomers, for different values of the drive, $E=0.2, 0.5$, and $1.0$, for the polyelectrolyte solution with trivalent counterions (see Fig. \ref{fig:Fig2}). (b) Variation of the fraction of monomers moving along the $+\hat z$ and $-\hat z$ directions, denoted by $P_{z+}$ and $P_{z-}$ respectively, with the external drive strength $E$. (c) The difference $P_{z-} - P_{z+}$ with electric field, obtained from the data in (b). (d) Variation of the average drift velocities of the monomers, $<v_{z+}>$ and $<v_{z-}>$, with electric field. The shaded regions in (b) and (d) denote the $E$-field range for the NDM regime.}

\label{fig:Fig3}
\end{figure}

In the following, we demonstrate numerically that the trapping mechanism described above, leads to the observed features: re-entrance, nonmonotonic $\mu \sim E$ dependence, and the anomalous NDM regime. Here we treat the monomer as a `test particle' (or a `tracer'/`tagged' particle) which is drifting through an environment of oppositely moving counterions that act as traps/obstacles. The motion of the counterions can be described and analyzed identically by considering them as the test particle and the monomers as traps.

First, we compute the distribution of the drift velocity $v_z$ of all the monomers, denoted by $P(v_z)$ and normalized to unity, for different values of the external drive. Three $P(v_z)$ distributions for external drives $E < E_m$, $E = 0.5 \approx  E_m$, and $E > E_m$ are shown in Fig. \ref{fig:Fig3}a. From this distribution we compute the fraction of monomers that are moving along the $+\hat z$ and the $-\hat z$ directions, denoted by $P_{z+} = \int_0^\infty P(v_z) dv_z$ and $P_{z-} = \int_{-\infty}^0 P(v_z) dv_z = 1-P_{z+}$ respectively. We also numerically compute the average drift velocities of these two groups, denoted by $<v_{z+}> = \int_0^\infty v_z P(v_z) dv_z$ and $<v_{z-}> = \int_{-\infty}^0 v_z P(v_z) dv_z$ respectively. 
From Fig. \ref{fig:Fig3}b, we find that both $P_{z+}$ and $P_{z-}$ are non-monotonic functions of the external drive $E$, and their difference $(P_{z-} - P_{z+})$ is shown in Fig. \ref{fig:Fig3}c. Below $E_m \approx 0.5$, as we increase the external drive, $(P_{z-} - P_{z+})$ increases, implying that more and more monomers are moving along the $-\hat z$ direction in response to the external drive, as expected. However above $E=E_m$,  the scenario reverses. In this case, more and more monomers start to move toward the {\it wrong} direction ($+ \hat z$), being trapped and dragged along by the on-coming counterions, causing $(P_{z-} - P_{z+})$ to decrease (Fig. \ref{fig:Fig3}c). The nonmonotonicity of $(P_{z-} - P_{z+})$ with $E$, at $E \approx 0.5$, gets directly reflected in the nonmonotonic behavior of mobility $\mu$ with $E$ (Fig. \ref{fig:Fig2}b). 
%
% 
% This trapping and dragging by the oppositely moving charges explains the nonmonotonic behavior of $\mu$ with $E$ and leads to the reentrance phenomenon, seen in Fig. \ref{fig:Fig2}, at around the same value of the electric field $E \approx 0.5$, as a consequence of this nonmonotonicity of $(P_{z-} - P_{z+})$ with $E$. 

Furthermore, in the shaded region of Figs. \ref{fig:Fig3}b and \ref{fig:Fig3}d, for $0.8 \lesssim E \lesssim 1.5$,
%the drag  on the monomers overwhelms the electrokinetic force due to the external drive. Thus, in this regime, 
as the drive is increased, the fraction of monomers moving toward the wrong direction and their drift velocity, $P_{z+}$ and $<v_{z+}>$, are found to increase sharply, whereas, the other fraction $P_{z-}$ decreases sharply and $<v_{z-}>$ remains fairly constant (dashed horizontal line in Fig. \ref{fig:Fig3}d). This complicated, and somewhat anomalous, dependence of these four quantities on the external field is precisely what triggers the counterintuitive NDM regime for $0.8 \lesssim E \lesssim 1.5$, as seen in Fig. \ref{fig:Fig2}c. In fact, the trapping and dragging events mentioned here can be observed directly by visually inspecting the trajectories of individual beads in the solution, see Fig. S6 in \textcolor{blue}{Supporting Information}.

For even larger drives, $E \gtrsim 1.5$, $P_{z+}$ and $P_{z-}$ become fairly constant, and their drift velocities $<v_{z+}>$ and $<v_{z-}>$ again increase roughly at the same rate. The system comes out of the NDM regime, and for $1.5 \lesssim E \lesssim 2.0$ we obtain $\mu_d \approx 0$ (zero differential mobility), see Fig. \ref{fig:Fig2}c. At even higher drive $E > 2.0$, the drift velocity $<v_z>$ shows a tendency to increase feebly, and one obtains a PDM regime again (see Fig. S7 in the \textcolor{blue}{Supporting Information}). Thus, NDM is triggered due to a delicate competition between the trap-and-drag mechanism and the electrokinetic motion of the charges. The electric field plays two antagonistic roles simultaneously, albeit with varying degrees of effectiveness depending on the field strength: one one hand, it pushes the charges to move faster, and on the other, it increases trapping probability that slows them down. All the features discussed above are a consequence of the complex interplay between these two mutually conflicting traits of the external drive.

Note that the re-entrance regime starts at lower $E$ for the trivalent counterions due to their enhanced electrostatic interactions that make the trap-and-drag mechanism much more effective compared to monovalent counterions. Furthermore, for the range of the external drive simulated here, NDM can be seen in the polyelectrolyte solution with trivalent counterions, but is absent in the monovalent case (see Fig. S8 in the \textcolor{blue}{Supporting Information}). In both cases we have the same density $\rho = 0.25$, and there are fewer particles in the trivalent case ($N = 4000$) compared to the monovalent case ($N = 6000$). This clearly demonstrates that, it is the nature of interactions in a system, and not how `crowded' the environment is, that determines whether or not NDM will be triggered, as has also been argued recently \cite{baiesi2015role}. However, it is possible that NDM also emerges for monovalent counterions at extremely large drives not easily accessible in our simulation.
\begin{figure}[htb]
\centering
{\includegraphics[width=8.cm]{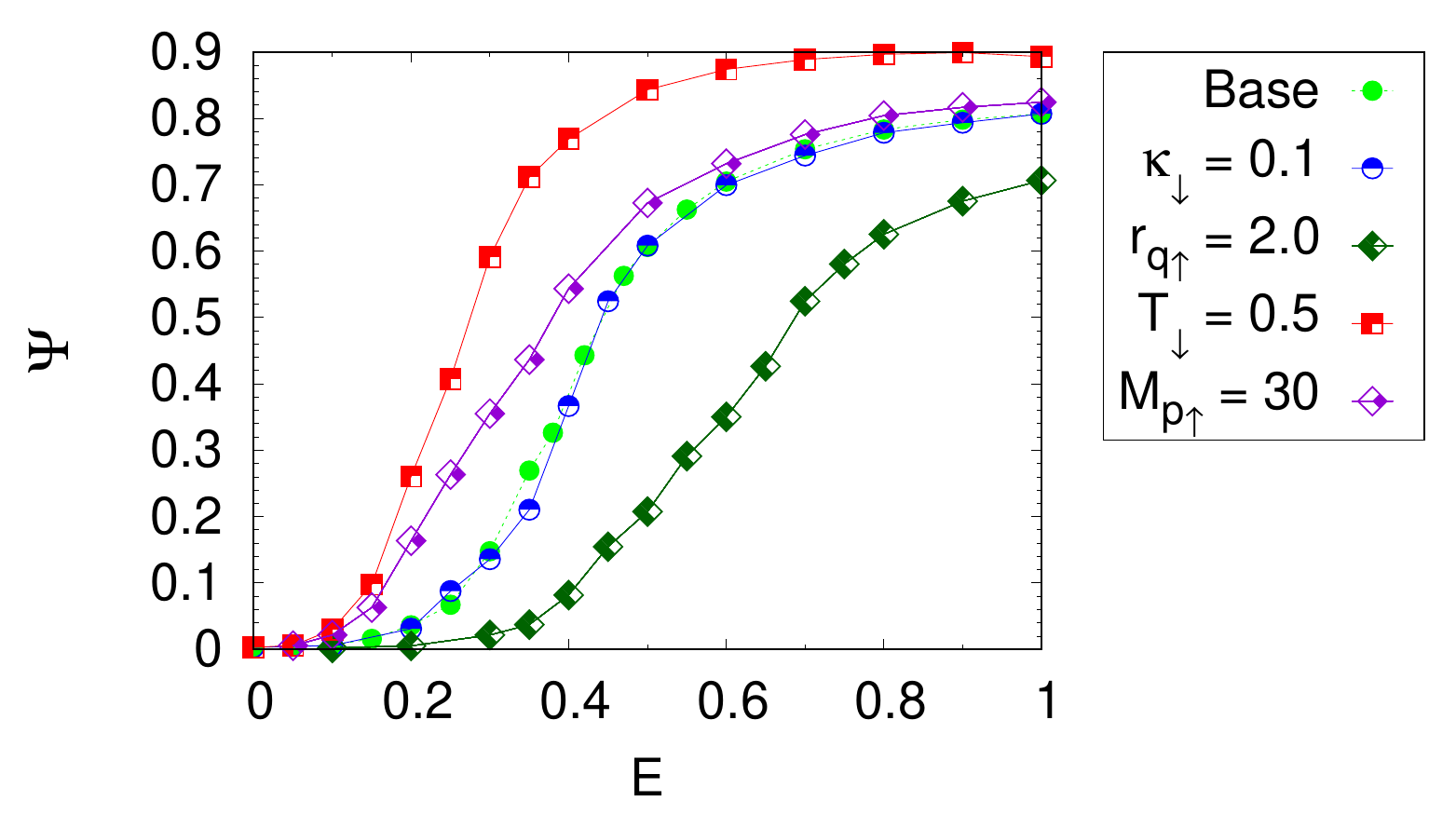}}
\caption{Dependence of the order parameter $\Psi(E)$ on different system parameters (for monovalent counterions). The reference curve, labeled as `Base', has the parameters $\kappa = 1$, $r_q = 2^{1/6}$, $T = 1$, and $M_p = 15$ (same as in Fig. 1). For each of the other curves, one of the above parameters has been altered, keeping all other parameters the same. The up ($\uparrow$) or down ($\downarrow$) arrows indicate whether the parameter has been increased or decreased, compared to the reference curve. For all the cases density $\rho = 0.25$ and $N = 6000$.}
\label{fig:Param}
\end{figure}

\paragraph{Dependence on system parameters.}
Finally, we systematically study the dependence of charge segregation on some of the important system parameters. The data are shown in Fig. \ref{fig:Param} where the curve labeled `Base' has the parameters $\kappa = 1$, $T = 1$, $r_q = 2^{1/6}$, and $M_p = 15$. Each of the other curves has one of these parameters altered, keeping all the others the same. We find that, for the electrostatic interactions chosen here, decreasing the Debye screening factor to $\kappa = 0.1$ has very little impact on the order parameter. This also seems to be in agreement with the results of 3D binary colloids at lower densities \cite{dzubiella2002lane}. However, strengthening electrostatics by increasing the cutoff from $r_q = 2^{1/6}$ to $r_q = 2$ shifts the phase transition toward higher external drives, as is expected: the monomers and the counterions are more strongly correlated in the latter case, and therefore one needs a stronger external drive for charge segregation.

At a lower temperature, $T = 0.5$, thermal noise competes weakly with the external drive, and thereby enhances the electrokinetic motion of the charged particles. This increases the friction between the oppositely moving charged particle fronts, and triggers charge segregation  at a lower external drive. Another way to think about this is in terms of the higher thermal diffusion of the charged particles at a higher $T$, which tends to keep the system homogeneous, whereas the external drive induces lane formation (uphill diffusion). Consequently, at a lower temperature it is easier to charge segregate the polyelectrolyte solution, and therefore the order parameter has a higher value for the same $E$-field.

\begin{figure}[htb]
\centering
{\includegraphics[width=5.5cm]{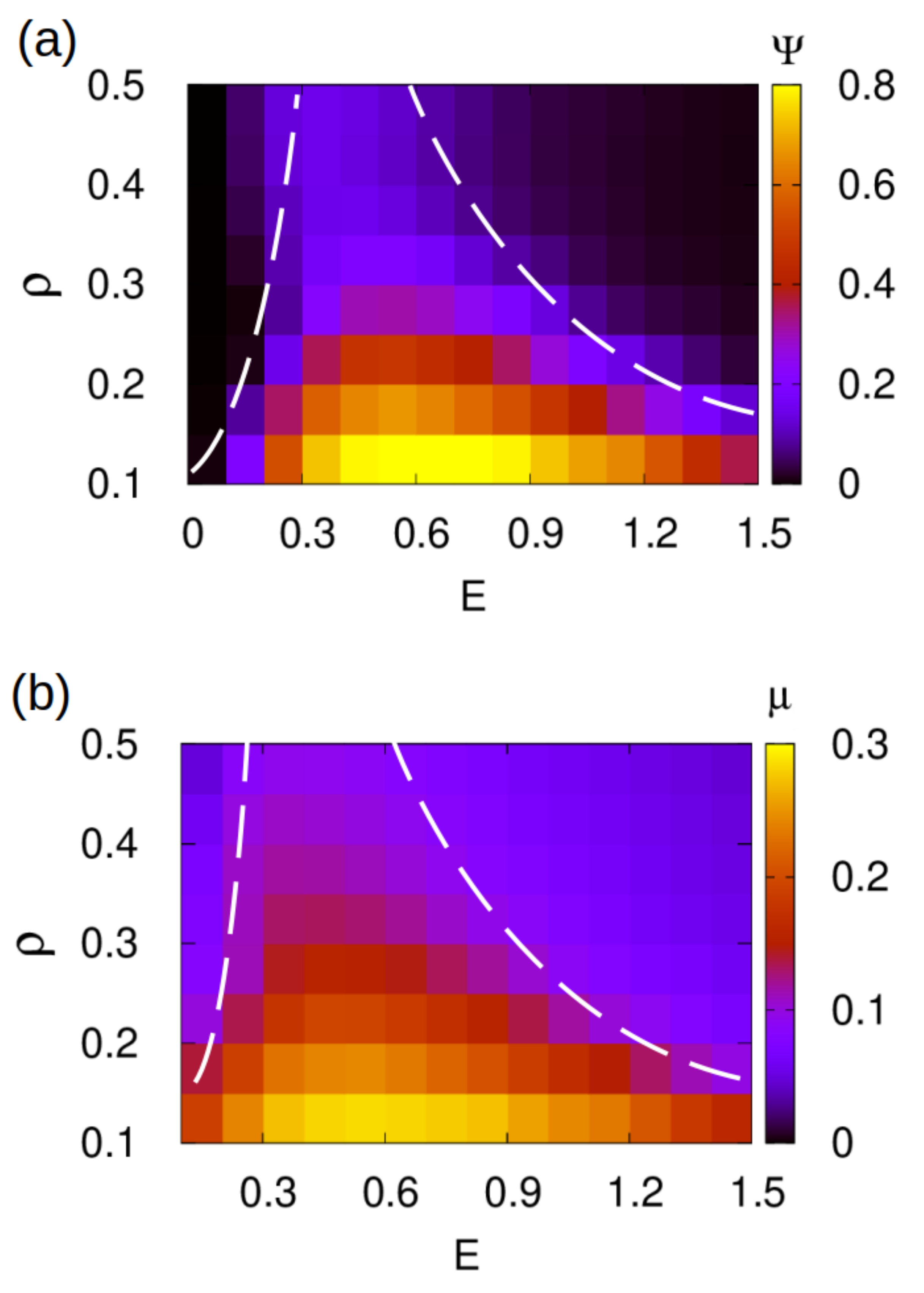}}
\caption{Phase diagrams in the $\rho-E$ plane depicting (a) the order parameter $\Psi$, and (b) the mobility $\mu$, for the trivalent case. The dashed lines are a guide to the eye. The area enclosed by the dashed lines roughly represents the region that exhibits prominent macroscopic charge segregation.}
\label{fig:PD}
\end{figure}

For a higher degree of polymerization $M_p = 30$, we find that the segregation starts at a lower $E$. When $M_p$ is higher, excluded volume interactions also become stronger, causing charge segregation at lower drives. Thus by changing these parameters as shown in Fig. \ref{fig:Param}, the threshold drive magnitude and the degree of phase segregation can be tuned as desired.

As is customary in studies of laning transition, we construct phase diagrams in the $\rho-E$ plane for the polyelectrolyte solution, indicating the charge segregation regime. The phase diagrams, showing the order parameter $\Psi$ and the mobility $\mu$, are presented in Fig. \ref{fig:PD} for polyelectrolytes with trivalent counterions (again, because of the computational ease, compared to monovalent counterions). The complete phase diagrams with monovalent counterions are expected to be qualitatively similar (see discussions in Figs. S9 and S10 of the \textcolor{blue}{Supporting Information}). It can be seen that the order parameter $\Psi$ is tightly correlated with the mobility $\mu$: the higher the mobility, the stronger is the phase segregation. From the phase diagrams, the segregation is found to be more pronounced for lower densities and is weaker for higher densities. This again can be explained from the fact that at high densities, electrokinetic motion of the charged particles is suppressed (see Fig. \ref{fig:PD}b), and consequently, the tendency toward phase segregation is lowered (Fig. \ref{fig:PD}a). Note that, even for high densities, there are distinct polyelectrolyte-rich and counterion-rich domains inside the simulation box in the presence of driving. However, for higher $\rho$, the polyelectrolyte-rich region gets infiltrated by more counterions and charge segregation is less pronounced, see Fig. S11 in \textcolor{blue}{Supporting Information} for related discussion. 
It is interesting to note that Fig. \ref{fig:PD}a is similar to the $\rho-E$ phase diagram for 2D rough disks under athermal conditions \cite{samsuzzaman2021reentrant}. This suggests that the properties for these driven soft-matter systems are quite generic, regardless of their specific details.

\section{Summary and outlook}
\label{conclusions}
To summarize, we have studied a coarse-grained polyelectrolyte solution model with explicit counterions in an implicit water-like solvent, driven by a constant external electric field. The interparticle electrostatic interaction is deliberately chosen to be a weak screened Coulomb potential, such that it can compete with the drive that need not be too large (see Fig. S12 in the Supporting Information for some data and discussion with full long-range Coulomb interactions). In the absence of the drive, the charged monomers and counterions form a homogeneously mixed phase ($\Psi \approx 0$). However, with the drive, the system spontaneously reorganizes, and forms two separate lanes of like-charges. This is a consequence of minimizing the energetically costly collisions (friction) between the oppositely moving monomers and counterions, thereby facilitating charge transport. Compared to charge symmetric and asymmetric electrolytes, the fully flexible polyelectrolyte chains have significantly stronger excluded volume interactions that lead to lane formation at rather low values of the external drive. This intriguing nonequilibrium phase transition exhibits no hysteresis and therefore has the signature of continuous equilibrium phase transition. At larger values of the external drive, we find that both the order parameter and the mobility decrease with the field, indicating a re-entrant transition. For the electric fields studied here, a complete re-entrance to a fully mixed phase is observed for the polyelectrolyte solution with trivalent counterions. The re-entrant phase is signaled by a sharp decrease of the mobility as the external drive is increased. Furthermore, we discover a regime where the differential mobility is negative in the presence of trivalent counterions, but such a regime is not observed for monovalent counterions under the same simulation conditions (for the $E$-field range investigated here). The re-entrance phenomenon, the decrease of the mobility with electric field, and the negative differential mobility are inter-related, and can be explained consistently by an intuitive trapping mechanism between the two species of charges, drifting opposite to each other in response to the external drive. Thus, our coarse-grained simulations unravel a rich and complex nonequilibrium behavior of driven polyelectrolyte solutions. It is pertinent to mention here that spontaneous phase segregation in the presence of a nonequilibrium external drive shares a few similarities with segregation phenomena reported in other kinds of nonequilibria, such as in the presence of activity; see for example  Refs. \cite{mccandlish2012spontaneous,stenhammar2015activity,smrek2017small,dolai2018phase} for phase separation in a mixture of active and passive (uncharged) particles.

We believe that the results presented above are the first steps toward a complete understanding of macroscopic charge segregation in driven polyelectrolytes. In the future, more detailed analyses can be performed to investigate various other aspects of charge segregation, re-entrance, negative differential mobility, and their dependencies on system parameters. Furthermore, these properties can be better understood by scrutinizing the conformations of the polyelectrolyte chains inside the lane, their velocity profiles perpendicular to the applied drive, the effects of geometric and dielectric confinement \cite{bagchi2020surface,bagchi2020dynamics}, asymmetric interactions and bead sizes, solvent quality, polyelectrolyte backbone rigidity, oscillatory external drives, and so forth. Moreover, in this work, hydrodynamic effects have been completely ignored, and it will be interesting to find out the extent to which the results, in different regions of the $\rho-E$ phase diagram, get modified by hydrodynamic interactions. We plan to focus on some of these issues in future works. The coarse-grained polyelectrolyte model studied here can be thought of as a representation of flexible polyelectrolytes, such as single-stranded DNA/RNA and sodium polystyrene sulfonate (NaPSS). Since the phase segregation in polyelectrolytes happens at much lower electric field drives compared to electrolytes (or colloids), it seems feasible to perform relatively simple laboratory experiments to verify the results presented in this work. Hopefully, improved understanding of the self-organization phenomena in such driven complex fluids will allow us to fabricate, control, and manipulate soft-matter systems more efficiently in the future.

\begin{acknowledgement}
The author thanks S. Dutta and U. Basu for useful discussions. The support of the Department of Atomic Energy, Government of India, under project no. {\bf RTI4001} is also acknowledged. The numerical computations were performed on the Mario HPC at ICTS-TIFR, Bengaluru, India. 
\end{acknowledgement}

%%%%%%%%%%%%%%%%%%%%%%%%%%%%%%%%%
% 
% \begin{suppinfo}
% The following files are available free of charge.
% \begin{itemize}
%   \item SI\_Lane.pdf: Contains additional figures and related discussions to better explain the results presented in this article.
% \end{itemize}
% \end{suppinfo}

%%%%%%%%%%%%%%%%%%%%%%%%%%%%%%%%%%%%%%%%%%

%\bibliographystyle{apsrev4-2}
%\bibliographystyle{unsrt}
\bibliography{References}

\vskip5cm

\begin{figure}[htb]
\centering
{\includegraphics[width=9cm]{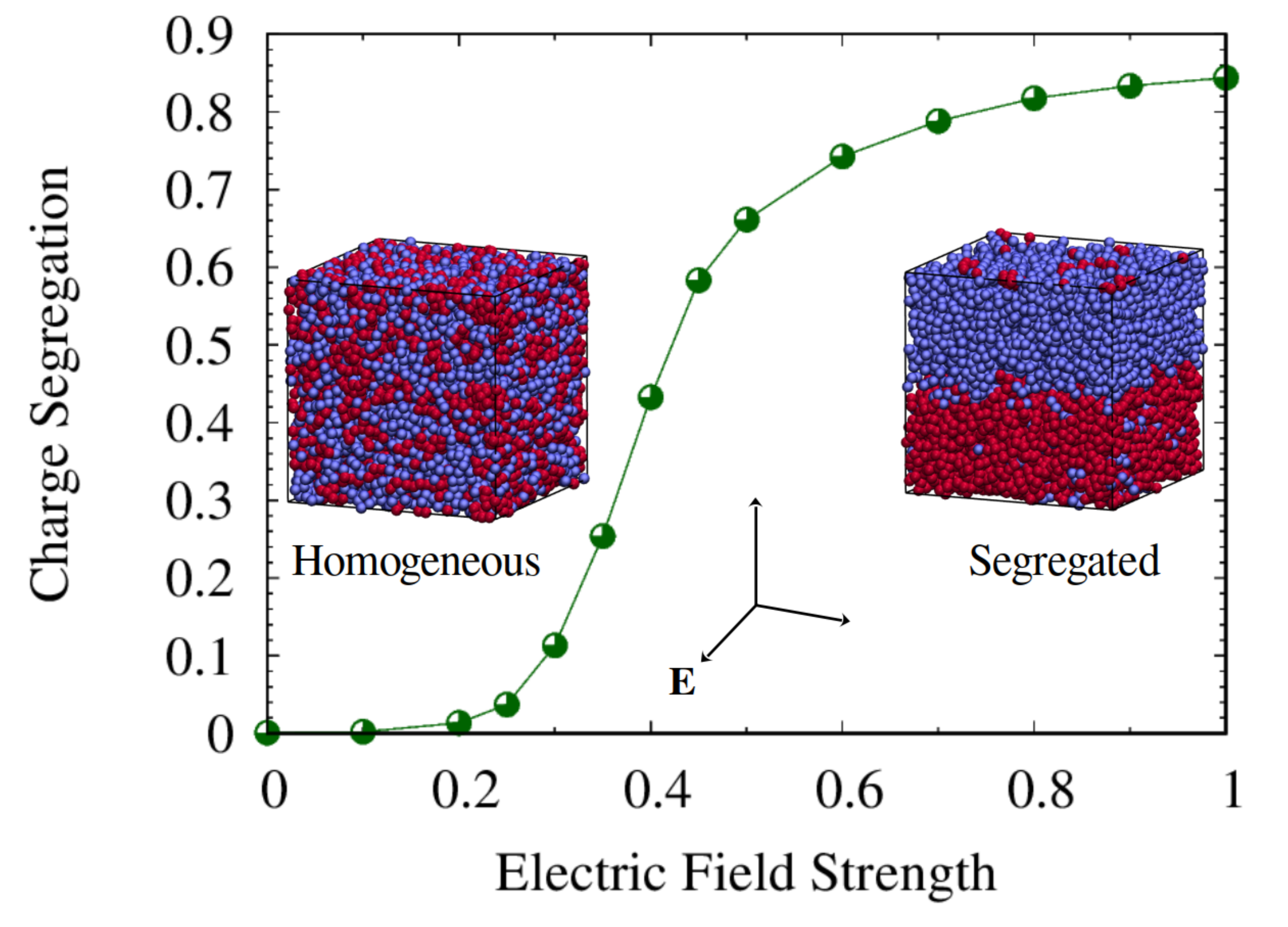}}
%\captionsetup{labelformat=empty}
\caption*{Graphical Table of Content Image}
\label{fig:TOC}
\end{figure}

%\onecolumngrid
\pagebreak
\section*{Macroscopic charge segregation in driven polyelectrolyte solutions (Supporting Information)}

%\centering
%\appendix

\renewcommand\thefigure{S\arabic{figure}}
\setcounter{figure}{0}    

%\vskip1cm

This document contains supporting figures and additional discussions to  help develop a clearer understanding of the charge segregation phenomenon in driven polyelectrolyte solutions.

\vskip4cm

\begin{figure}[htb]
\centering
{\includegraphics[width=8cm]{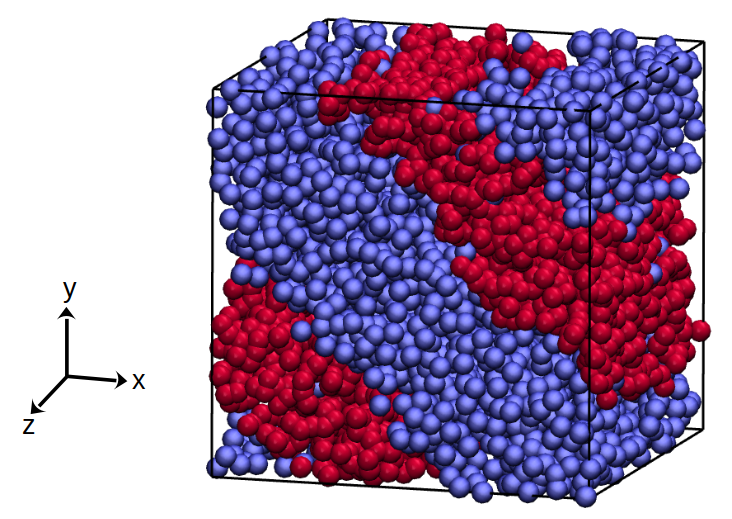}}%\hskip1cm
\caption{Snapshot of the polyelectrolyte solution for $E=0.9$ and $\rho=0.1$. Here the red beads represent the monomers and the blue beads represent the counterions (monovalent). Electric field is along the $+\hat z$ direction. In some cases, charge segregation is found to be tilted with respect to both $\hat x$ and $\hat y$ directions. Since the density is low here, tilted configurations with multiple interfaces, do not lead to a significant energy cost, thereby making such a configuration much more stable. It is possible that such a configuration is a long-lived meta-stable state, since inter-particle interactions are weak for low density. Nevertheless, the emergence of such long-lived tilted charge segregation is still quite remarkable. Such a charge configuration will be energetically costlier for higher $\rho$ (stronger interactions), and therefore, unlikely to be observed at late times. Also note that all these charge configurations form spontaneously, and consequently, it is entirely possible that, for all parameters kept the same, different initial conditions will lead to different charge configurations in the NESS, but with the same average value of the order parameter $\Psi$.}
\label{fig:tilted}
\end{figure}

\begin{figure}[htb]
\centering
{\includegraphics[width=14cm]{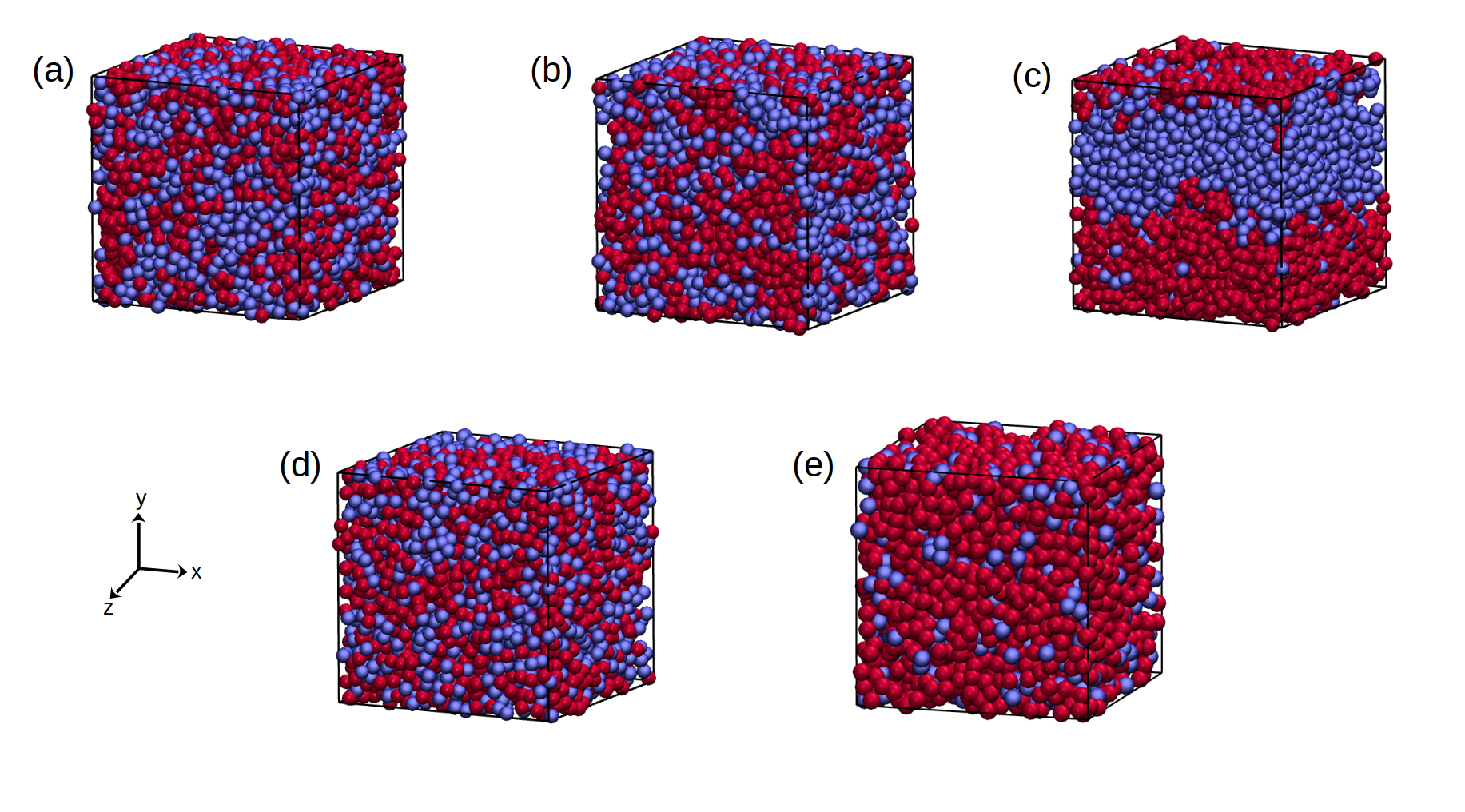}}%\hskip1cm
\caption{Snapshots for different values of $M_p$ for $E = 1.0$: (a) $M_p = 2$ (b) $M_p = 3$ (c) $M_p = 4$, (d) 1:1 electrolyte ($M_p = 1$), and (e) 1:4 electrolyte ($M_p = 1$). Electric field is along the $+\hat z$ direction.  Only (c) shows a clear charge segregation. None of the rest show any discernible structure in their charge distribution, implying a homogeneously mixed phase, $\Psi \approx 0$ (no segregation); see Fig. 1h in main text.}
\label{fig:Mp}
\end{figure}

\begin{figure}[htb]
\centering
{\includegraphics[width=15cm]{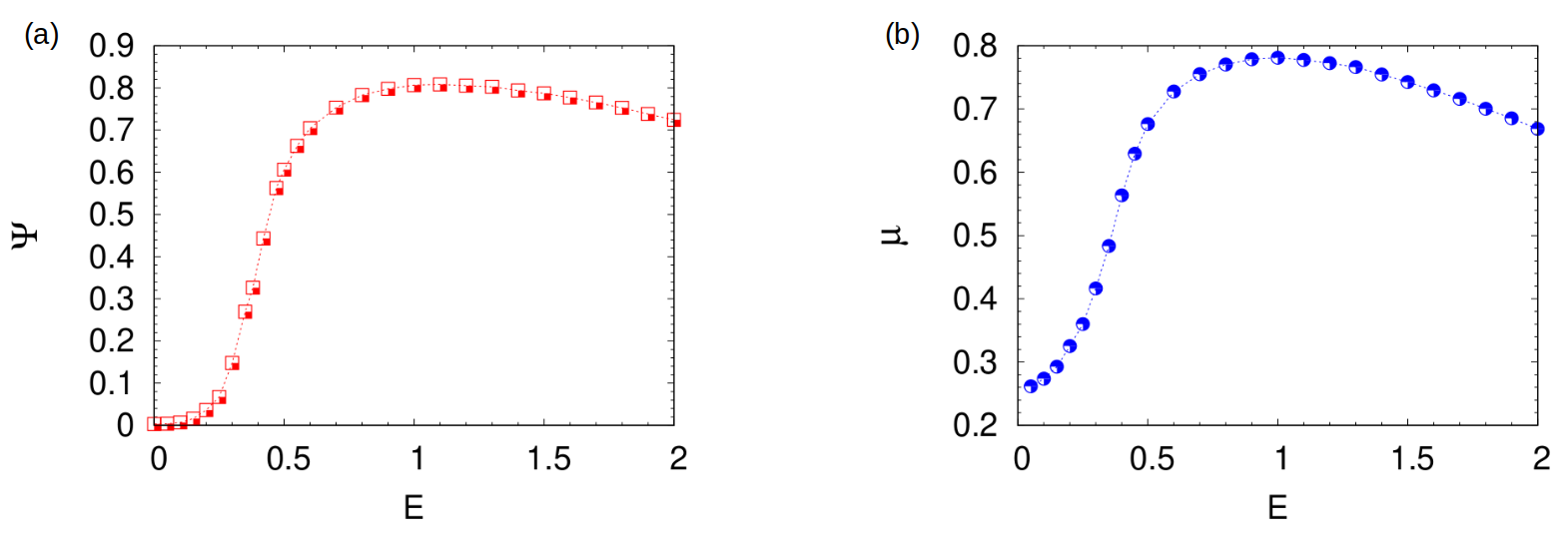}}\hskip1cm
\caption{Variation of (a) order parameter $\Psi$, and (b) mobility $\mu = \mu_p = \mu_c$ with the external drive in the range $0 \leq E \leq 2$. For $E \gtrsim 1.0$, we find that $\Psi$ and $\mu$ decrease with increasing external drive. Here $N = 6000$ and $\rho = 0.25$. }
\label{fig:OPMonoFull}
\end{figure}

\vskip0.5cm
\begin{figure}[htb]
\centering
{\includegraphics[width=10cm]{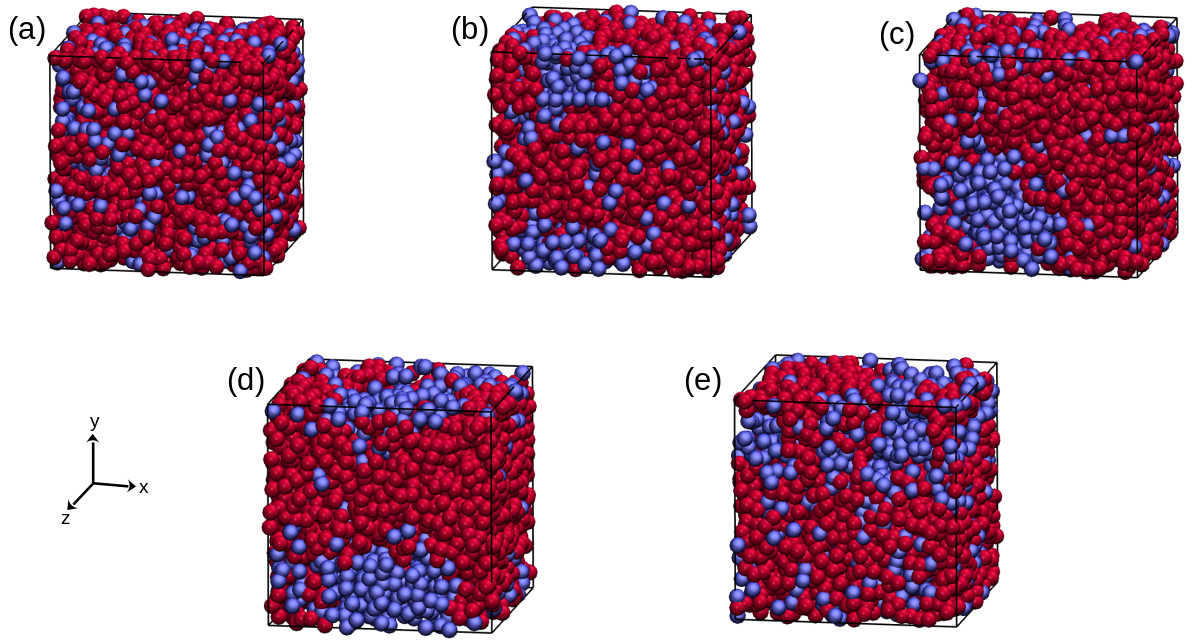}}%\hskip1cm
\caption{Snapshots of the polyelectrolyte solution with trivalent counterions for different values of the external electric field: (a) E = 0.1, (b) E = 0.2, (c) E = 0.5, (d) E = 0.8, and (e) E = 1.2. Here the red beads represent the monomers and the blue beads represent the trivalent counterions. Note that here the number of counterions $N_c = \frac 13 N_p \times M_p$. Electric field is along the $+\hat z$ direction. Charge segregation is weaker here compared to a polyelectrolyte solution with monovalent counterions.}
\label{fig:TriConfigs}
\end{figure}

\begin{figure}[htb]
\centering
{\includegraphics[width=9cm]{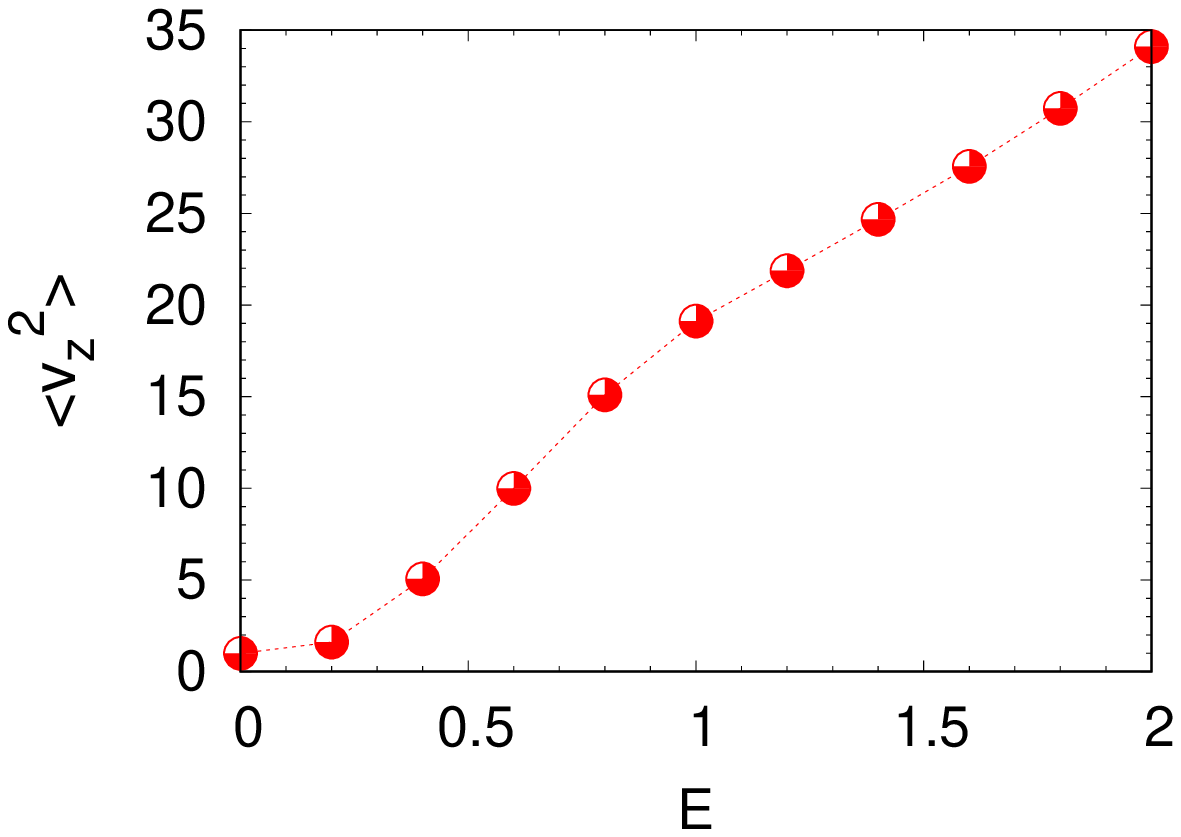}}%\hskip1cm
\caption{Variation of $<v^2_z>$ for the polyelectrolyte chains with the applied electric field $E$, for the case with trivalent counterions. In equilibrium ($E = 0$), we have $<v^2_z> = T$ following {\it energy equipartition theorem}, where $T = 1$ is the temperature of the solution set by the Langevin thermostat. Notice that it is not apparent from this data that a regime with NDM exists for $0.8 \lesssim E \lesssim 1.5$ in this case.}
\label{fig:v2E}
\end{figure}

\begin{figure}[htb]
\centering
{\includegraphics[width=17cm]{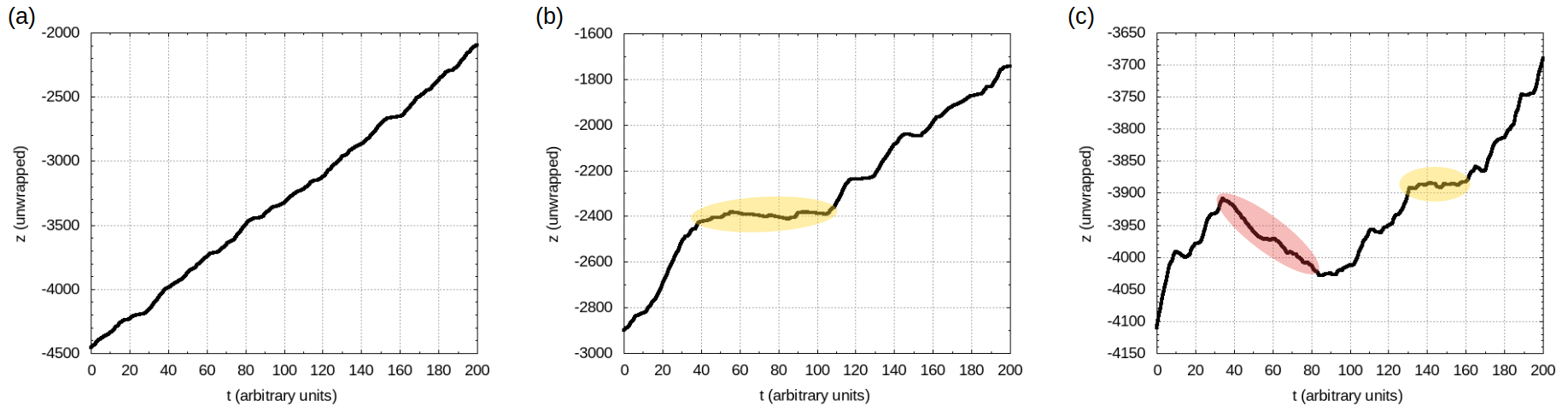}}%\hskip1cm
\caption{The z-coordinate (unwrapped) of a trivalent counterion in the polyelectrolyte solution is shown, for the electric field value $E = 1.0$ which falls inside the NDM regime (see Fig. S7 below). We choose any time instant in the NESS and label it as $t = 0$ (the origin of time is irrelevant in the NESS). Each unit spacing along the time axis (x-axis) corresponds to $1000$ iterations of the Langevin dynamics simulation (with time-step $\Delta t = 0.001$). The z-position is measured in units of $\sigma$. In (a), (b), and (c), we trace the trajectories of three counterions for the same time window, $0 \leq t \leq 200$. In (a) is shown the typical trajectory of a counterion that is moving fairly unhindered in response to the applied electric field ($z-$coordinate increases roughly monotonically). In (b), we find that the counterion is majorly stalled for a considerable length of time, $40 \lesssim t \lesssim 100$ (no change in its z-position: yellow shaded region). This indicates the trapping by the opposite species, since there are no other mechanisms that can cause stalling, in the presence of a constant $E$-field drive. In (c) the counterion starts to move in the {\it wrong} direction (toward $-\hat z$ direction: red shaded region) even with the external drive switched on, for an appreciable time span, $30 \lesssim t \lesssim 80$, and for a distance $\approx 125 \sigma$. This shows the effect of the {\it trap-and-drag} mechanism produced by the oppositely moving charges. All these features will also be present for any monomer in the solution.}
\label{fig:Trajectory}
\end{figure}

\begin{figure}[htb]
\centering
{\includegraphics[width=6cm]{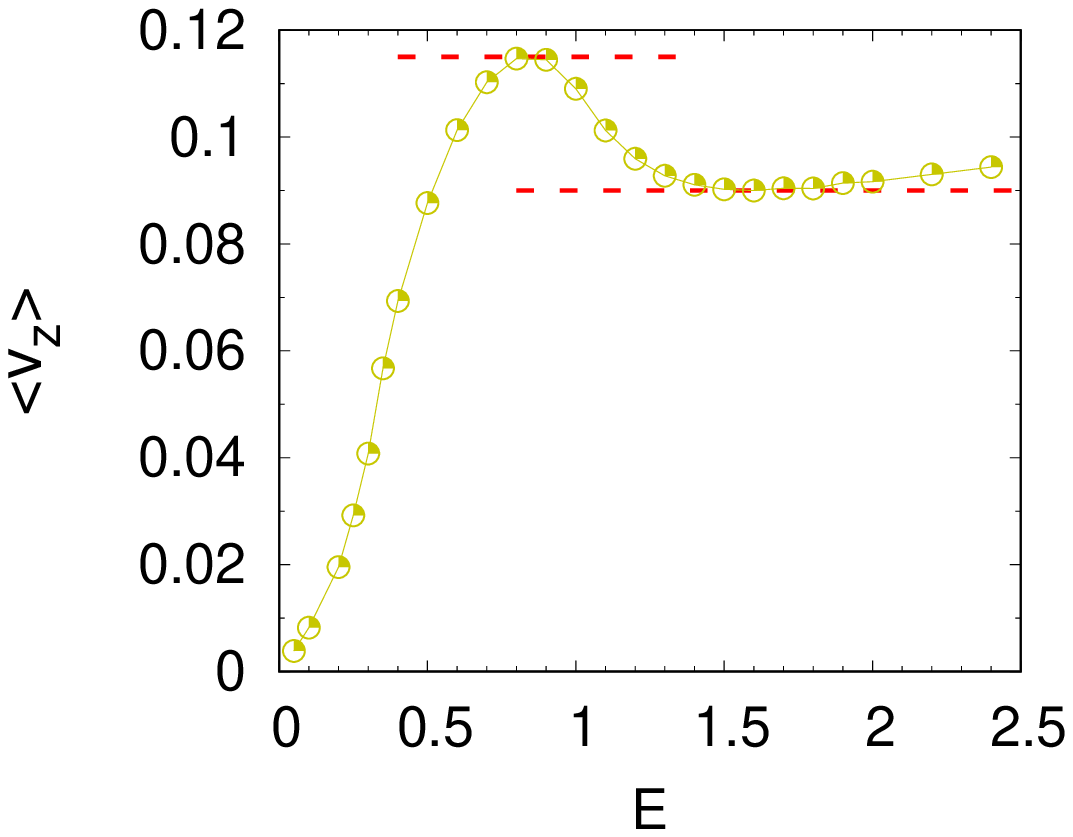}}%\hskip1cm
\caption{Variation of the $<v_{z}>$ for the polyelectrolyte solution with trivalent counterions beyond $E = 2.0$. Here, $<v_z> = -\frac 1q <v_{zp}> = \frac 1{3q} <v_{zc}>$ (see main text). As the external drive is increased from zero, the polyelectrolyte solution exhibits transitions from PDM $\rightarrow$ NDM $\rightarrow$ PDM. There is also a fairly extended regime of ZDM (zero differential mobility), for $1.5 \lesssim E \lesssim 2.0$, between the NDM and the PDM regime. The dashed lines are a guide to the eye. Note that the existence of the re-entrance regime at $E = 0.5$ (Figs. 3a and 3b in main text) is not apparent from this data.}
\label{fig:MobV}
\end{figure}

\begin{figure}[htb]
\centering
{\includegraphics[width=6cm]{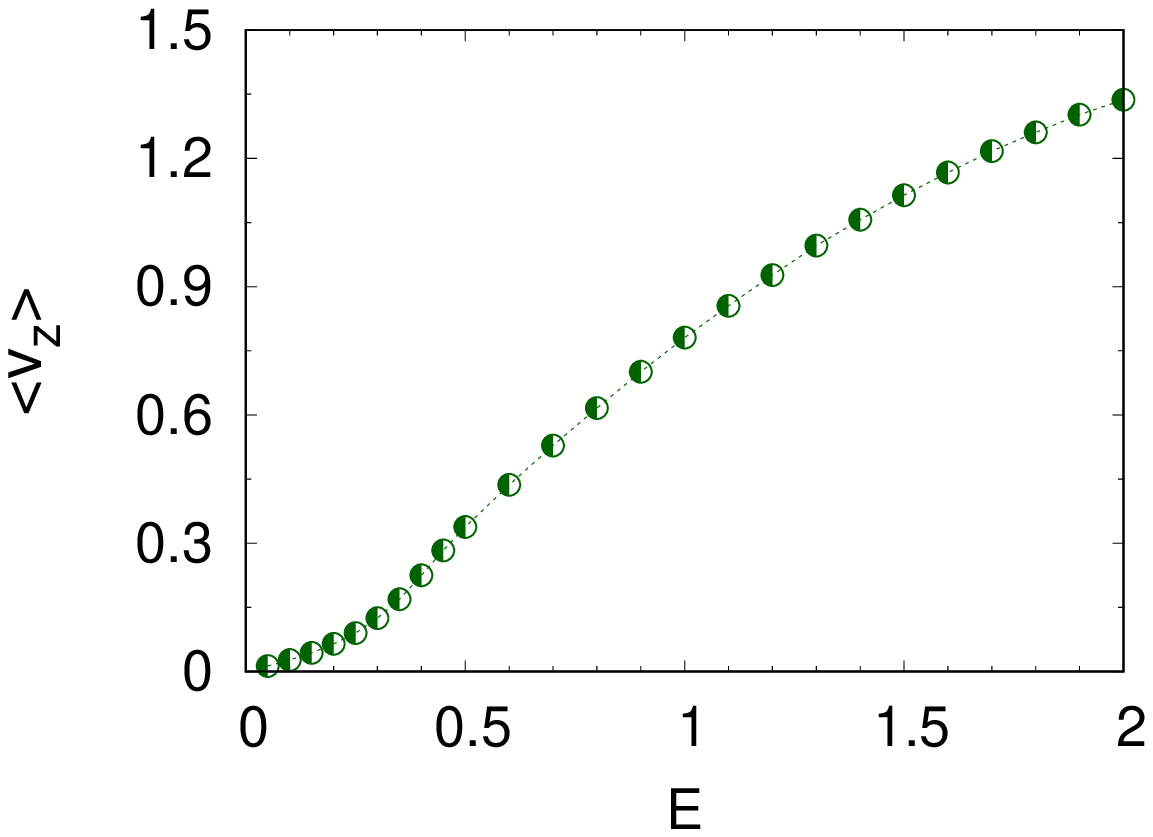}}%\hskip1cm
\caption{Variation of the $<v_{z}>$ for the polyelectrolyte solution with monovalent counterions. Here, $<v_z> = -\frac 1q <v_{zp}> = \frac 1{q} <v_{zc}>$ (see main text). For this case, no NDM regime is observed in the range $0 \leq E \leq 2$, unlike trivalent monomers. However, it is possible that NDM emerges for monovalent counterions at very large external drives.}
\label{fig:Monovz}
\end{figure}

\begin{figure}[htb]
\centering
{\includegraphics[width=6cm]{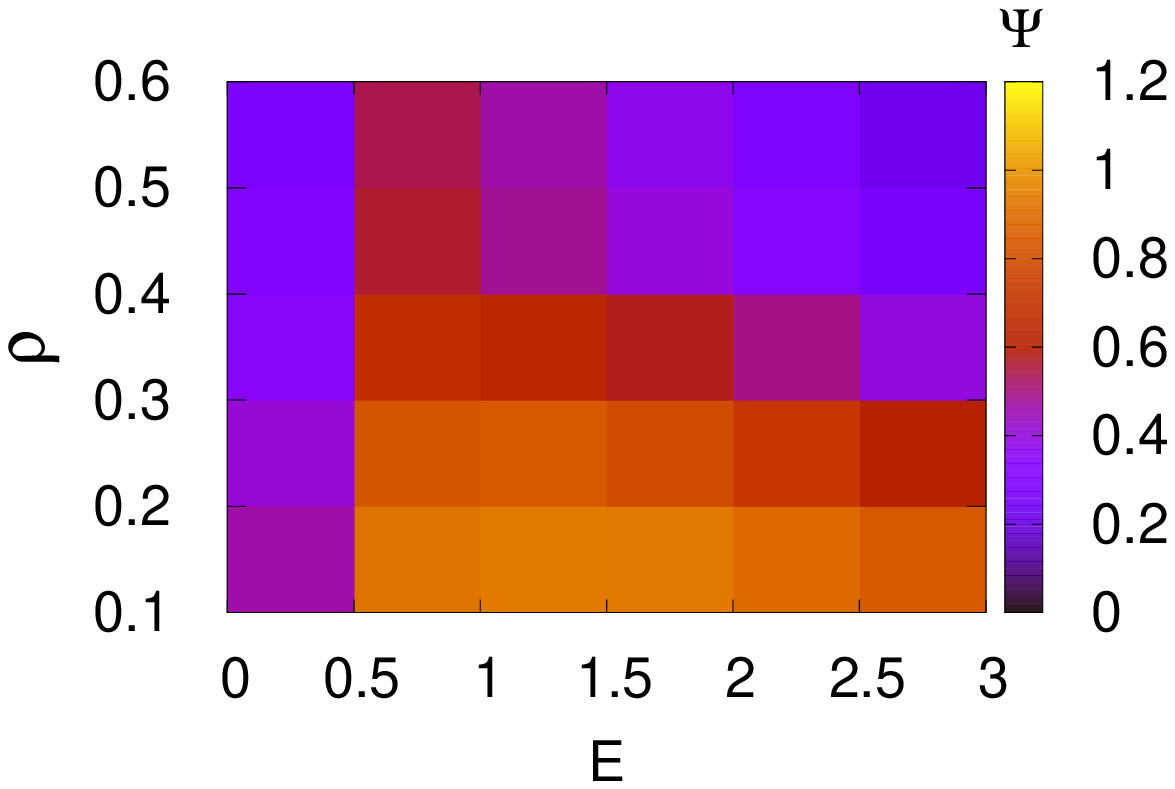}}\hskip1cm
{\includegraphics[width=6cm]{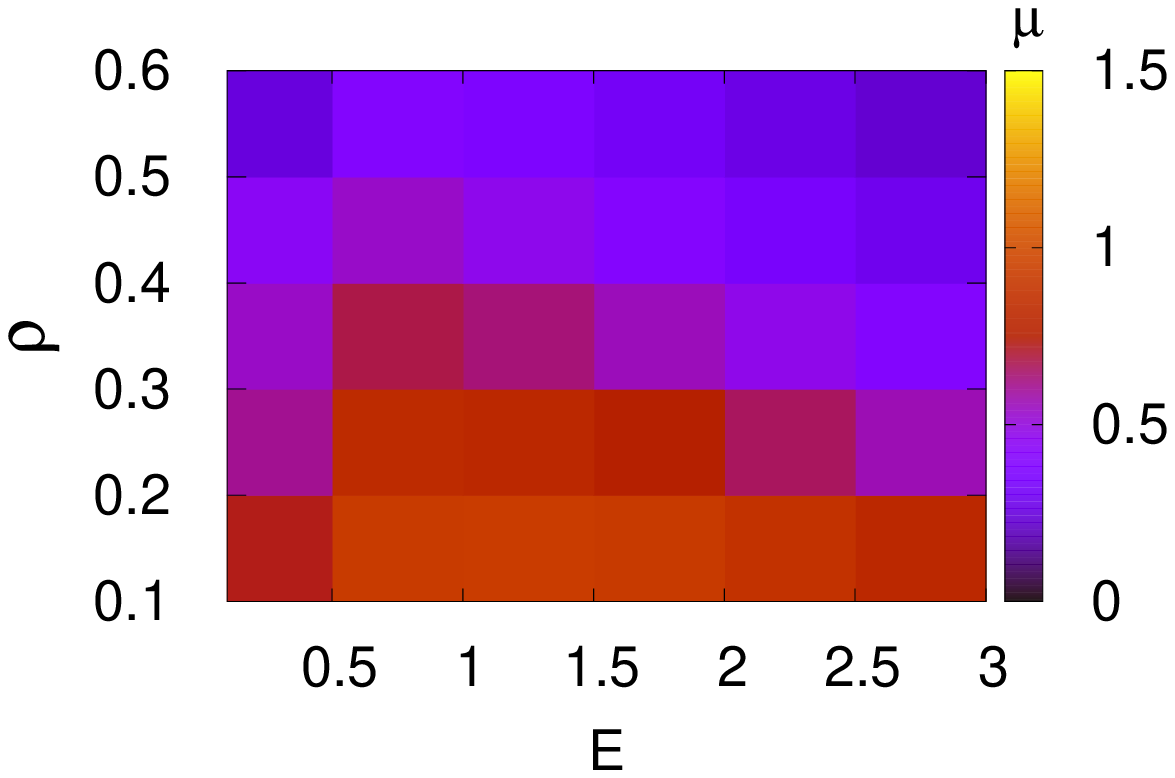}}\hskip1cm
\caption{Phase diagrams for the polyelectrolyte solution with monovalent counterions, for temperature slightly different than the phase diagrams in the main text (we have chosen $T = 0.5$ here for computational convenience). All the other parameters remain the same. For the monovalent case one needs to go to higher external drives to get a complete picture of its phase behavior. Nonetheless, all phase diagrams have qualitatively similar features (see Fig. 6 in main text).}
\label{fig:PDsds}
\end{figure}

\begin{figure*}[htb]
\centering
{\includegraphics[width=14cm]{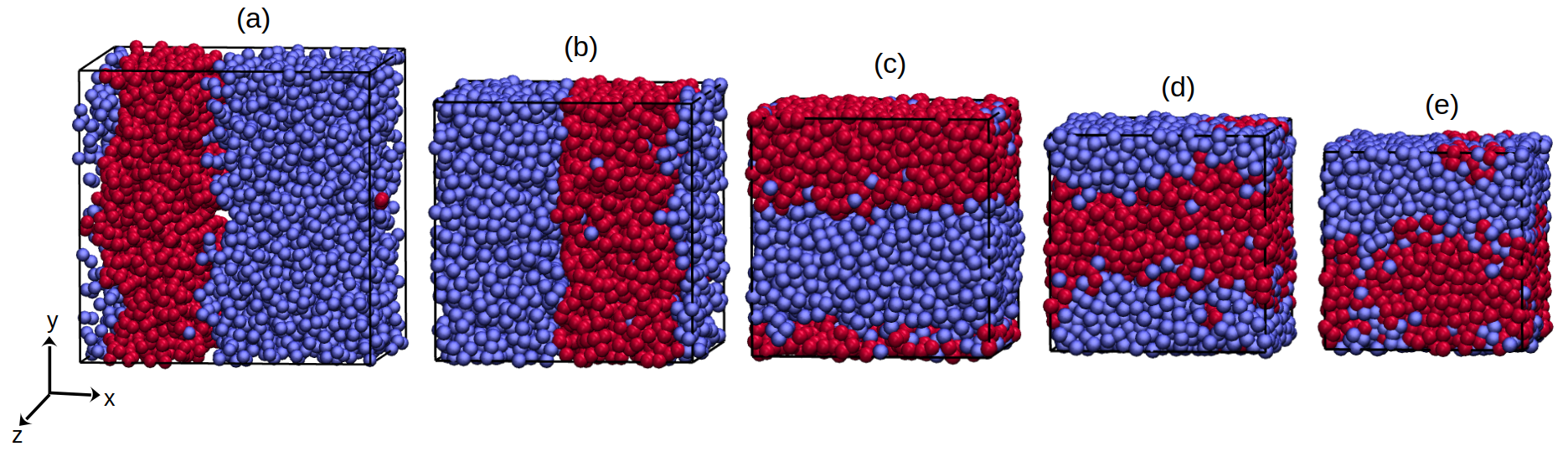}}%\hskip1cm
\caption{The snapshot of the simulation box for a few different densities: (a) $\rho = 0.1$, (b) $\rho = 0.2$, (c) $\rho = 0.3$, (d) $\rho = 0.4$, and (e) $\rho = 0.5$ showing the charge segregation. The monomers are represented by red beads and the (monovalent) counterions are represented by blue beads. Here, $E = 1.0$ and $N = 6000$. Electric field is along the $+\hat z$ direction.}
\label{fig:AllRho}
\end{figure*}

\begin{figure}[htb]
%\centering
{\includegraphics[width=12cm]{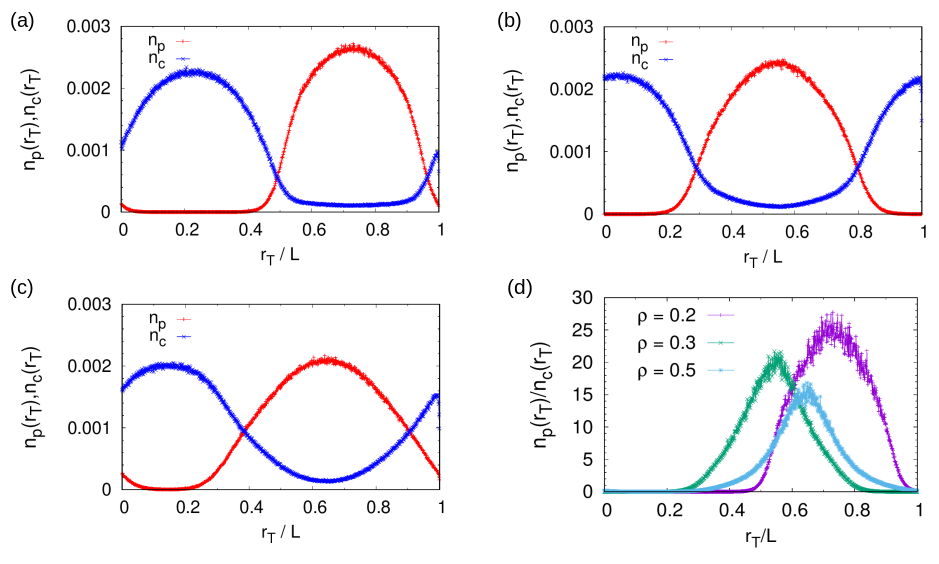}}\hskip1cm
\caption{Number density profiles for the polyelectrolyte monomers $n_p$ and monovalent counterions $n_c$, along the directions transverse to the $E$-field, denoted by $r_T \equiv {\hat x,\hat y}$, for (a) $\rho = 0.2$, (b) $\rho = 0.3$, and (c) $\rho = 0.5$. Distinct polyelectrolyte-rich and monomer-rich regions can be seen clearly in all the cases (see Fig. S10 above). In (d) we show the ratio $n_p/n_c$, a low value of which is an indicator of the higher presence of counterions in the polyelectrolyte-rich region. The ratio is more spatially diffuse and has a lower peak value for higher $\rho$. This makes charge segregation less pronounced for higher $\rho$, as seen in the $\rho-E$ phase diagrams. Here the electric field is $E = 1.0$  and $N = 6000$ for all the cases.}
\label{fig:DP}
\end{figure}

\begin{figure}[htb]
\centering
{\includegraphics[width=17cm]{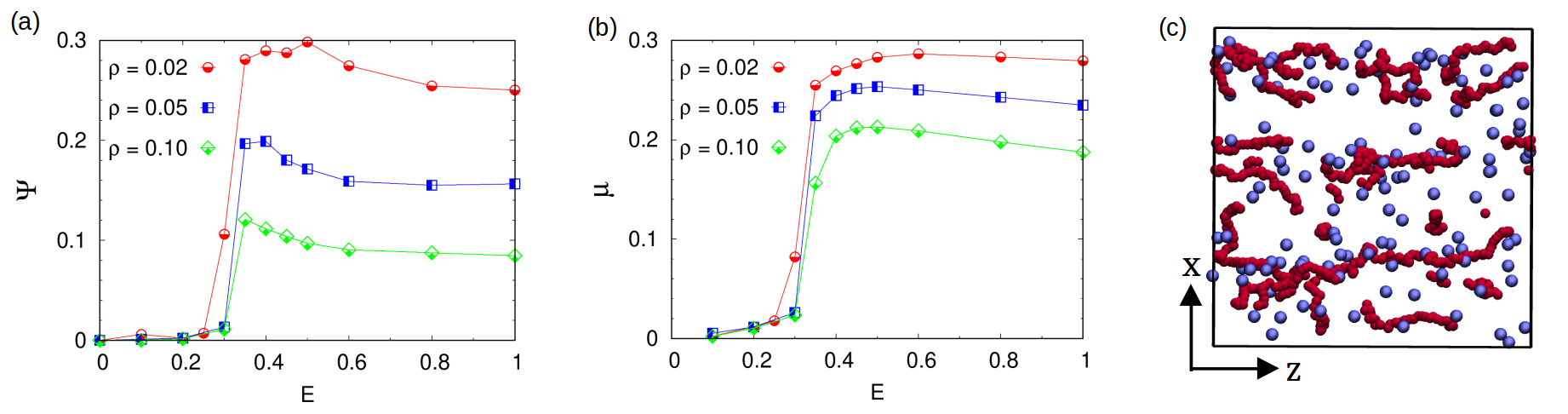}}\hskip1cm
\caption{(a) The order parameter $\Psi(E)$ is shown for a polyelectrolyte solution with trivalent counterions in the presence of {\it long-range Coulomb interaction}, $V_q(r) = k_BT {\ell}_B / r$. The long-range Coulomb interactions have been computed with the {\it particle-particle particle-mesh} (PPPM) method with an accuracy of $10^{-4}$. The simulation parameters used here are $N = 1000$, $M_p = 30$, $N_p = 25$, and $T = 0.5$. 
% 
% Following Figs. 5 and 6 of the main text, we have chosen a higher degree of polymerization $M_p$, lower temperature $T$, and very low densities $\rho \leq 0.1$, so as to see charge segregation well within the computationally convenient range, $0 < E < 1$, even with long-range Coulomb interactions. 
% 
The equilibrium phase ($E = 0$) in this case is a clustered phase that is broken apart when the electric field drive is switched on. The order parameter has been calculated in the same way as for screened interactions. The variation of $\Psi(E)$ shows an interesting nonmonotonic behavior with the external drive, suggesting that some degree of charge segregation happens even with long-ranged Coulomb interaction when the polyelectrolyte solution is driven beyond a threshold electric field strength. As expected, charge segregation in this case is much poorer compared to screened Coulomb interactions. For the electric fields simulated, we have not found two macroscopic lanes of like charges, but multiple lanes seem to form, just as in binary colloidal mixtures with screened Coulomb interactions. The data also display a hint of a re-entrance behavior (nonmonotonic $\Psi(E)$ at higher $E$). (b) The corresponding mobility $\mu = \mu_p = \mu_c$ is shown as a function of the drive $E$ for different densities. The variation of $\mu(E)$ is also nonmonotonic. A typical simulation snapshot, for $\rho = 0.02$ and $E=0.5$, is depicted in (c), showing a slice of the x-z plane. To develop a proper understanding of the nonequilibrium properties of driven polyelectrolytes with long-range Coulomb interactions, and how they differ from polyelectrolytes with screened Coulomb interactions, more in-depth analyses need to be performed.}
\label{fig:LRC}
\end{figure}
% 
% \pagebreak
 
% \begin{tocentry}
% \includegraphics[width=8cm]{TOC.pdf}
% \end{tocentry}
% 
\end{document}